\documentclass[acmsmall,nonacm]{acmart}
\AtBeginDocument{%
  \providecommand\BibTeX{{%
    \normalfont B\kern-0.5em{\scshape i\kern-0.25em b}\kern-0.8em\TeX}}}

\setcopyright{acmcopyright}
\copyrightyear{2018}
\acmYear{2018}
\acmDOI{XXXXXXX.XXXXXXX}

\acmConference[Conference acronym 'XX]{Make sure to enter the correct
  conference title from your rights confirmation emai}{June 03--05,
  2018}{Woodstock, NY}
\acmBooktitle{Woodstock '18: ACM Symposium on Neural Gaze Detection,
 June 03--05, 2018, Woodstock, NY} 
\acmPrice{15.00}
\acmISBN{978-1-4503-XXXX-X/18/06}

\usepackage{xcolor}
\usepackage{colortbl} 
\usepackage{physics}
\usepackage{amsmath}
\usepackage{tikz}
\usepackage{mathdots}
\usepackage{cancel}
\usepackage{color}
\usepackage{siunitx}
\usepackage{array}
\usepackage{multirow}
\usepackage{gensymb}
\usepackage{tabularx}
\usepackage{extarrows}
\usepackage{booktabs}
\usetikzlibrary{fadings}
\usetikzlibrary{patterns}
\usetikzlibrary{shadows.blur}
\usetikzlibrary{shapes}
\usepackage{caption}
\usepackage{subcaption}
\usepackage{float}
\usepackage{changepage}
\usepackage{soul}
\sethlcolor{green}
\usepackage{marginnote}
\usepackage[colorinlistoftodos]{todonotes}
\usepackage{listings}

\usepackage{graphicx,lipsum,wrapfig}

\usepackage{multicol}

\usepackage{tikz}
\newcommand{\myarrow}[1][0.1pt]{\tikz[baseline=-0.26em,y=3em, x=3em]{\filldraw[line width=#1] (0.4202,0.0021) .. controls (0.4202,-0.0000) and (0.4188,-0.0018) .. (0.4171,-0.0025) .. controls (0.3917,-0.0092) and (0.3699,-0.0236) .. (0.3509,-0.0401) .. controls (0.3355,-0.0538) and (0.3225,-0.0704) .. (0.3130,-0.0890) .. controls (0.3119,-0.0915) and (0.3094,-0.0929) .. (0.3066,-0.0929) .. controls (0.3028,-0.0929) and (0.2996,-0.0897) .. (0.2996,-0.0858) .. controls (0.2996,-0.0848) and (0.3000,-0.0837) .. (0.3003,-0.0827) .. controls (0.3087,-0.0665) and (0.3193,-0.0517) .. (0.3316,-0.0391) -- (0.1181,-0.0391) .. controls (0.1143,-0.0391) and (0.1111,-0.0359) .. (0.1111,-0.0320) .. controls (0.1111,-0.0282) and (0.1143,-0.0250) .. (0.1181,-0.0250) -- (0.3471,-0.0250) .. controls (0.3611,-0.0137) and (0.3766,-0.0046) .. (0.3935,0.0021) .. controls (0.3766,0.0088) and (0.3611,0.0179) .. (0.3471,0.0292) -- (0.1181,0.0292) .. controls (0.1143,0.0292) and (0.1111,0.0323) .. (0.1111,0.0362) .. controls (0.1111,0.0401) and (0.1143,0.0432) .. (0.1181,0.0432) -- (0.3316,0.0432) .. controls (0.3193,0.0559) and (0.3087,0.0707) .. (0.3003,0.0868) .. controls (0.3000,0.0879) and (0.2996,0.0889) .. (0.2996,0.0900) .. controls (0.2996,0.0939) and (0.3028,0.0970) .. (0.3066,0.0970) .. controls (0.3094,0.0970) and (0.3119,0.0956) .. (0.3130,0.0932) .. controls (0.3225,0.0745) and (0.3355,0.0580) .. (0.3509,0.0443) .. controls (0.3699,0.0278) and (0.3917,0.0133) .. (0.4171,0.0067) .. controls (0.4188,0.0059) and (0.4202,0.0042) .. (0.4202,0.0021) -- cycle;}}

\begin{document}

\title[Recognising misinformation based on retracted science]{Enable people to identify science news based on retracted articles on social media}




\author{Waheeb Yaqub}
\affiliation{%
  \institution{The University of Sydney - School of Computer Science}
  \country{Australia}
}
\email{waheeb.faizmohammad@sydney.edu.au}

\author{Judy Kay}
\affiliation{%
  \institution{The University of Sydney - School of Computer Science}
  \country{Australia}
}
\email{judy.kay@sydney.edu.au}

\author{Micah Goldwater}
\affiliation{%
  \institution{The University of Sydney - School of Psychology}
  \country{Australia}
}
\email{micah.goldwater@sydney.edu.au}

\renewcommand{\shortauthors}{Yaqub, et al.}

\begin{abstract}

For many people, social media is an important way to consume news on important topics like health. Unfortunately, some influential health news is misinformation because it is based on retracted scientific work. Ours is the first work to explore how people can understand this form of misinformation and how an augmented social media interface can enable them to make use of information about retraction. We report a between subjects think-aloud study with 44 participants, where the experimental group used our augmented interface. Our results indicate that this helped them consider retraction when judging the credibility of news. Our key contributions are foundational insights for tackling the problem, revealing the interplay between people's understanding of scientific retraction, their prior beliefs about a topic, and the way they use a social media interface that provides access to retraction information.
\end{abstract}

\ccsdesc[500]{Human-centered computing~Social content sharing}
\ccsdesc[500]{Human-centered computing~Social media}
\ccsdesc[500]{Human-centered computing~Social networks}
\ccsdesc[500]{Human-centered computing~Social networking sites}
\ccsdesc[300]{Human-centered computing~User centered design}

\keywords{retraction, misinformation, social media, science news}

\begin{teaserfigure}
  \includegraphics[width=\textwidth]{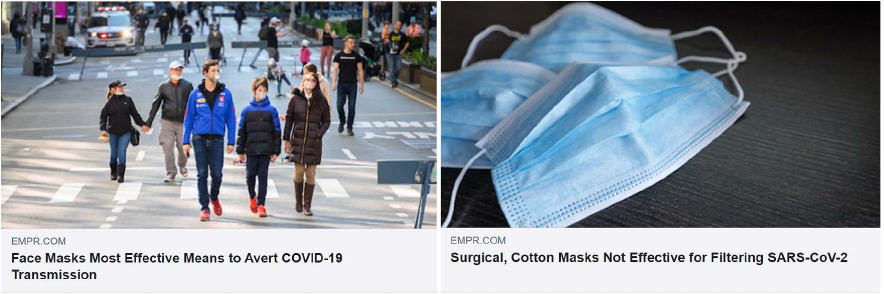}
  \caption{Retracted studies? Same news source and same health topic, but opposite claims. Which one can we trust? Our work explores how to help people realise that the right one was based on a retracted scientific paper.}
  \Description{Enjoying the baseball game from the third-base
  seats. Ichiro Suzuki preparing to bat.}
  \label{fig:teaser}
\end{teaserfigure}


\maketitle

\section{Introduction}
\label{sec:introduction}

 \color{black}Many people rely on social media for their news about important aspects of their lives, such as health
~\cite{hitlin2018science,funk2017how}.
There are infamous cases where social media has spread misinformation based on scientific publications that have been retracted. 
For example, the paper reporting a link between serving bowl size and food consumption\color{red}~\cite{wansink2005super}\footnote{\color{black}We used red font to distinguish normal references from retracted ones and added the word RETRACTED in the title of a paper as it would generally appear on publisher webpages.}\color{black}.
Five years after retraction, some people who have thousands of followers and scientific credentials continue to propagate its scientific claim in social media posts\footnote{The statistics can be accessed through Altmetric Explorer \textcolor{blue}{\href{https://www.altmetric.com/explorer/highlights?q=\%20Bowls\%3A\%20Serving\%20Bowl\%20Size\%20and\%20Food\%20Consumption&scope=all&show_details=101749553}{platform.}}}.
\color{black} According to the guidelines of Committee on Publication Ethics (COPE), retraction constitutes a formal withdrawal of a published scientific article due to misconduct or errors that may invalidate the study’s conclusions and claims~\cite{cope2019retraction}. \color{black}
Misinformation that is based on retracted science is an under-studied contributor to the systematic dissemination of disinformation and misinformation\footnote{By definition, misinformation is incorrect or misleading information presented as facts~\cite{Misinfor1:online}. Where this is intentional, it is called disinformation.
 By this definition, any person who learned about science findings that were later retracted is a misinformed person.}~\cite{serghiou2021media}.
Worryingly, this form of misinformation has been increasing over time~\cite{brainard2018whatAM,cokol2008retraction}.
\color{black}



There is growing understanding about the extent of retraction of scientific papers.
This shows a steady increase in the number of retractions~\cite{cokol2008retraction}.
For example, the RetractionWatch database increased 10-fold over 10 years~\cite{brainard2018whatAM}, currently standing at 32k retracted articles. 
Even the recent topic of COVID-19 had 72 retracted articles in 2020,
rising to 205 articles in 2021~\cite{oransky2021review}.  
In 2018, retracted articles made up 0.04\% of those published~\cite{brainard2018whatAM}.
However, this may be the tip of the iceberg, as the methods for detecting problem papers are limited~\cite{bordino2020retracted,alfirevic2020retracted}. 
A 2020 study~\cite{bik2016prevalence,gopalakrishna2022prevalence}
points to problems in up to 12.4\% of published articles due to use of inappropriate image duplication
\cite{bik2016prevalence}.
Such work also highlights failures in reviewing processes. 
Moreover, in the last two decades, over 12,000 journals have not recorded a single retraction as per Web of Science database~\cite{brainard2018whatAM}.
Misinformation created in this ecosystem affects diverse stakeholders~\cite{yaqub2020bias}, 
including experts and scientists~\cite{kataoka2022retracted}, practitioners~\cite{steen2011retractions}, science communicators (news media), policymakers~\cite{salinas2020altmetrics} and especially laypeople. \color{black}We focus on laypeople on social media for the following reasons: First, laypeople are farther away from the source in terms of expertise and reach, and they trust scientists and the scientific community~\cite{funk2020science}. Second, reported science news plays a crucial role in shaping people's understanding of science~\cite{hargreaves2003towards} and news is heavily consumed through social media~\cite{walker2021news} and also science news~\cite{abhari2022twitter}. Third, pre-retraction online media attention far outweighs post-retraction attention of scientific studies~\cite{serghiou2021media} and there are no syncing tools like CrossMark~\cite{wager2015why}, Zotero, Mendeley to inform social media users about such retractions.\color{black}

We particularly focus on \textit{health news} that people consume on social media platforms \color{black} because it is so important.
This is due to its wide dissemination, consumption and influence on important decisions that people make.
Health news is widely disseminated on news media that ends up on social media ~\cite{ladheri2016nutrition}.
Many people rely on social media to get science news~\cite{hitlin2018science} on important topics.
For example, Figure~\ref{fig:teaser} shows examples of news about mask wearing and COVID-19 \color{black}that were disseminated from published scientific articles. \color{black}
Together, these factors mean that science news that is \color{black} based on retracted scientific papers
can misinform a large number of people~\cite{abhari2022twitter}.
People then draw on such news to make important decisions about health and wellness.
\color{black} 


\color{black}
\subsection*{Research Questions}

One way to tackle this problem is with better interfaces for reading social media news.
Essentially, we envisage augmenting current SMNP interfaces so that people can take account of retraction-based misinformation.
Our work aims to provide the foundations for understanding how this can be done effectively.
This involves are two important challenges for people.
First, people need to \textit{understand the notion of scientific retraction} and how retraction affects the credibility of the claims in the retracted publication.
To gain insights about this, we aim to answer the following two research questions: 
\color{black}
\begin{itemize}
    \item \textbf{RQ1}: How do people understand the retraction of scientific publications?
    \item \textbf{RQ2}: What reasons for retractions do people consider to impact credibility of scientific claims?
\end{itemize}
\color{black}
The first question aimed to discover what overall understanding people have about scientific retraction.
The second then digs into this concept in terms of seven major reasons for retractions. 
Three reasons, fabrication, falsification and plagiarism account for about half of all retractions
~\cite{brainard2018whatAM} $-$
they fall under the US government definition of scientific misconduct~\cite{brainard2018whatAM} and are also listed in the U.S. Office of Research Integrity ~\cite{research2021integrity}.
Non-fraudulent retractions represent around 40\% of retractions~\cite{oransky2020retraction, brainard2018whatAM}.
The four such reasons we consider are
\textit{errors}, \textit{reproducibility}, \textit{permission}, \textit{duplicate publication} and \textit{plagiarism}.

Our study was designed to start with exploration of these two questions.
By the end of this part of the study, all participants should have had some understanding of retraction and the reasons for it. 
They had also considered the impact of these on credibility of the scientific claims.
This was a foundation for exploring how people could make use of information about retraction.
\color{black}

\color{black}
\color{black}

Once aware of retraction, people currently face a second challenge in learning when a social media news post (SMNP) is based on retracted science. 
We envisage addressing this by augmenting SMNP interfaces with retraction information 
using a light-weight approach similar to the 
Fakey~\cite{avram2020exposure} Fact Check button.
We wanted to understand how people would make use of such information.
To do this we explore:
\begin{itemize}

  \color{black} \item \textbf{RQ3}: How does the availability of retraction information interact with people’s prior beliefs when they rate the credibility of SMNPs? \color{black}

     \item \textbf{RQ4}: How does reading SMNP with retraction information change people’s belief on health topics?
\end{itemize}


\color{black}

The motivation for RQ3 is to establish a basis for an interface design that accounts for differences in people's prior beliefs~\cite{pennycook2021psyschology,shahid2022matches,ecker2020can}.
RQ4 aims to see the impact of available retraction information.



To answer these questions, we designed a \color{black} 
think-aloud study with 44 participants with both qualitative and quantitative analyses (mixed methods), \color{black} all psychology students at the University of Anon.
The first part explored their understanding of retraction and provided a foundation tutorial about it (RQ 1 and 2).
Participants then rated 12 social media news posts on three health topics.
The 22 participants in the Control Group used an interface similar to popular social media sites.
The other 22 Treatment Group participants had one additional button on their interface, enabling them to see information about retraction.
We assessed the prior and post beliefs about the topics.
All the study materials and research questions were pre-registered on the Open Science Foundation (OSF) project  \textcolor{blue}{\href{https://osf.io/pt9fe/?view_only=0c76301e5d0b47148fda69b8f6ccc8a8}{online}}. Our data will also be available under that OSF project, once the study is published. 

Next we describe the background with related works, study design, then the results, and finally discuss the implications of the results for the research literature and design of social media interfaces. 



\section{Background and Related Work}
\label{sec:relatedwork}
This section first provides background on the state of the art in fake news, particularly when it is based on retracted information.
Then it reviews work on the way people assess credibility of information.
We show how our work fills a gap in the literature since there has been no work that has explored the challenges of understanding how to help people to recognise social media misinformation that is based on retracted scientific publications.


\subsection*{Fake News}
False information can be classified as misinformation or disinformation according to the intent of the generator and propagator of the false information \cite{wardle2017information}. 
Allcott et al. \cite{allcott2017social} provides a widely agreed definition of fake news: `news articles that are intentionally and verifiably false and could mislead readers'. 
This is disinformation, the \textit{intentional} spread of false information, especially to sway public opinion on a topic or for materialistic gains~\cite{allcott2017social,mirza2023tactics}.
By contrast, misinformation may not be intentional and it can occur in many ways;
for our work, an important case is where reporters create science news that is based on retracted publications.
\subsection*{Retraction}
Previous work has used the terms, correction and retraction interchangeably to indicate an update to false-labelled information ~\cite{walter2020metaanalytic,stubenvoll2021whyretractions}.
In our context they have different impact.
According to the COPE 2019 guidelines~\cite{cope2019retraction},  
retraction is not appropriate when the main findings of the study are still reliable and correction can solve errors in the published study. Retraction should occur when findings are deemed unreliable~\cite{cope2019retraction}. 

\color{black}
\subsection*{Factors Affecting Credibility}
We now discuss the notion of credibility and the factors that influence people to believe information~\cite{fogg1999elements}.
Early work by Fogg~\cite{fogg1999elements} identifies many factors at play in credibility assessment of an interface.
These include familiarity with topic, look of the interface, similarity to the author, labels of expertise, reputation of the source, personal experience, and authority which Fogg et al. extended in a large-scale study on the credibility of health websites~\cite{fogg2003howdo}.  
These ideas are useful also for the case of social media posts, as in recent work by other researchers in the field, in both quantitative ~\cite{jahanbakhsh2022leveraging,bhuiyan2020investigating} and qualitative~\cite{hadlington2022perceptions} research.\color{black}


A user's prior knowledge, cognitive biases, and personal traits play a pivotal role in how they determine the credibility of news posts and articles~\cite{kim2021systematic}. Shahid et al.~\cite{shahid2022matches} noted that having prior knowledge of the topic helps individuals assess the credibility of fake multimedia content. Another significant factor is the identity of the person who posted or shared the article; posts from friends are perceived as more credible than those from individuals like politicians, yet less credible than content shared by experts~\cite{karlsen2021social}.

Tandoc et al.~\cite{tandoc2018defining} emphasized that social media users consume news from a wide range of sources. Additionally, the platform on which the news is shared, such as Reddit versus Twitter, and the reputation of the news outlet, also contribute to credibility perceptions. Organizations that are less known are generally considered to be less reputable~\cite{allcott2017social}. 

Furthermore, social engagement statistics, including the number of likes, shares, and comments on platforms like Facebook, have a substantial impact. A study conducted in 2020~\cite{avram2020exposure} found that these statistics strongly influenced users to interact with articles of lower credibility. Consequently, higher engagement statistics often result in users engaging with and disseminating low-credibility articles within their social networks without verifying their accuracy~\cite{geeng2020fake,koch2020effects}. Tandoc et al. (2018) coined this phenomenon as a "self-fulfilling cycle," which allows low-credibility posts to accumulate increasing engagement through the bandwagon effect.

We were unable to find work that has studied how people can take account of retraction information when judging the credibility of social media news.
However, we can draw on broader work on correcting misinformation.
For example, a 2021 study by Stubenvoll and Mathess~\cite{stubenvoll2021whyretractions} highlights the challenges in effectively communicating correction to headlines that involve numbers.
The work highlighted that numerical retraction fails because participants are anchored to initial number. 
O`Rear and Radvansky~\cite{o2020failure} found that retracted information is used even after its retraction, so creating misinformation after its retraction. 
Notably, Ecker et al. ~\cite{ecker2020can} suggest that repeating novel misinformation in corrections does not enhance misinformation, therefore making it safe to do. 
Work on retracted political news by~\cite{stubenvoll2021whyretractions, o2020failure, ecker2020can} involves very different considerations from those in the science publication process with different implications for credibility assessment. 

\color{black}
The only study that uses retraction in a similar context to our study is by Greitemeyer~\cite{greitemeyer2014article}. It examined if people still believed in a retracted article's findings after they have been informed about the retraction. 
The work is based on a paper that reported a causal link between a person's elevated height and increased real prosocial behaviour\color{red}~\cite{sanna2011rising}\color{black}. \color{black} 
Participants were allocated to three groups:  
a debriefing group, a no-debriefing group, or a control group. 
The debriefing group was told that the published research article they learned about was retracted due to fabricated data. This was printed in bold and a large font. Participants in this group were also informed that there was no scientific evidence for the study's claim.  
The no-debriefing group had no information about the retraction, and the control group did not learn about the article at all. The study explored if people changed their beliefs after hearing about the retraction. 
The results showed that participants in the debriefing group were less likely to believe the findings compared to the no-debriefing group $p-value = .043$. However, the debriefing group still had a higher chance of believing in the findings compared to the control group.$p-value = .045$. These findings suggest that a retraction note on the published article is not sufficient to ensure that readers of the published article no longer believe in the study's conclusions.  The work did not involve social media news posts.


\subsection*{Lightweight interventions in credibility assessments}

Unlike Greitemeyer’s study~\cite{greitemeyer2014article}, our work involves enhancing retraction information on a social media interface, as opposed to modifying the actual published article page. Numerous researchers have experimented with different warning labels to encourage users to evaluate news posts critically. Clayton et al.~\cite{clayton2020real} discovered that fake news accompanied by generic warnings were perceived as less accurate by social media users; similar outcomes were observed by Koch et al.~\cite{koch2020effects} and Yaqub et al.~\cite{yaqub2020effects}. Building upon this notion, Gwebu et al.~\cite{gwebu2021can} found that specific warnings pertaining to news posts had greater efficacy than general warnings on social media platforms. The presence of fact-check tags or labels on news posts also influenced perceived credibility~\cite{clayton2020real}. Sharevski et al.~\cite{sharevski2022meaningful} noted that many participants desired contextual information in conjunction with a warning red flag, serving as a nudge to recognize and evade misinformation. Similarly, aligned with providing contextual information, Jahanbakhsh et al.~\cite{jahanbakhsh2021exploring} investigated nudges prompting users to rationalize the accuracy of a claim, resulting in a reduction in the dissemination of false news. Additionally, Bhuiyan et al.~\cite{bhuiyan2021nudgecred} extended the concept of red flags, introducing positive flags in their nudging messages by categorizing posts as Reliable, Questionable, or Unreliable.

\color{black}
All these approaches have complex effects on people's behaviour.
For example, as a consequence of such interventions, participants engaged less with fake news.
Yaqub et.al~\cite{yaqub2020effects} and Lees et al.~\cite{lees2021twitter} found that sharing intent was reduced when SMNP was presented with a warning label.
This was more so for politically left-leaning people than those who were right-leaning. 
Inadvertently, 
such interventions can also reduce sharing of true news~\cite{jahanbakhsh2021exploring}. 
Further, Pennycook et al~\cite{pennycook2020implied} found that people judged unlabelled true news as more accurate and therefore shared it more while Yaqub et al. ~\cite{yaqub2020effects} reported no significant impact of credibility warning labels on true news sharing.  
Finally, 
Pennycook et al. found that veracity of the headline had no impact on sharing intentions~\cite{pennycook2021shifting}.
\color{black}

\color{black}
\subsection*{Our Contribution}
There have been many studies based on warnings labels for social post features.
These embedded new elements in the SMNP interface to help users assess news credibility. 
Our work builds upon that approach; 
we propose a new interface element for retraction,
to enable better informed science and health news consumption. \color{black} 


Our work also builds on the large body of work on analyzing and understanding the social and cognitive mechanisms contributing to the exponential spread of false news ~\cite{zhou2019fake, mosleh2021perverse, pennycook2019fighting}. 
Previous research has identified cognitive mechanisms that can underpin the design of interfaces and warning labels to combat the spread \color{black}of such pre-existing beliefs~\cite{yaqub2020effects}, social crowd influence ~\cite{micallef2020role}, politically motivated reasoning~\cite{pennycook2021psyschology,pennycook2018lazy} and users' attention~\cite{momen2021nudgecred}. 
Literature described above on misinformation has been dominated by large-scale quantitative studies and on US crowd-source samples ~\cite{allcott2017social,shrestha2021analysis,lees2021twitter,sherman2020designing,epstein2021explanations,allen2021scaling,jahanbakhsh2021exploring,koch2020effects,pennycook2019fighting,ecker2020can,clayton2020real,yaqub2020effects,sharevski2022meaningful,stubenvoll2021whyretractions}.
Our work aims to complement that approach with a detailed think-aloud study that can provide a rich picture of people's understanding and thinking.
This should be valuable to build better social media interfaces.

This work breaks new ground as it is the first study of news based on retractions of scientific papers.
The nature of scientific publication makes this quite different from the work on fake news in political domains~\cite{garrett2013thepromise, stubenvoll2021whyretractions,walter2020metaanalytic} $-$
political sources which are quite different from scientific papers~\cite{peng2021dynamics}. \color{black}
This is because the scientific process underpins the validity of scientific papers and the scientific community is more trusted~\cite{funk2020science}. 
Some work has studied misrepresentation and retraction of scientific studies in the news media~\cite{dumas2017poor} and on social media~\cite{serghiou2021media}. 
But ours is the first work to explore how people \textit{understand retraction} and how interfaces could help people account for it in assessing the credibility of science news on social media.



\section{Method}
\label{sec:method}

\begin{figure*}
    \centering
    \includegraphics[width=1.0\textwidth]{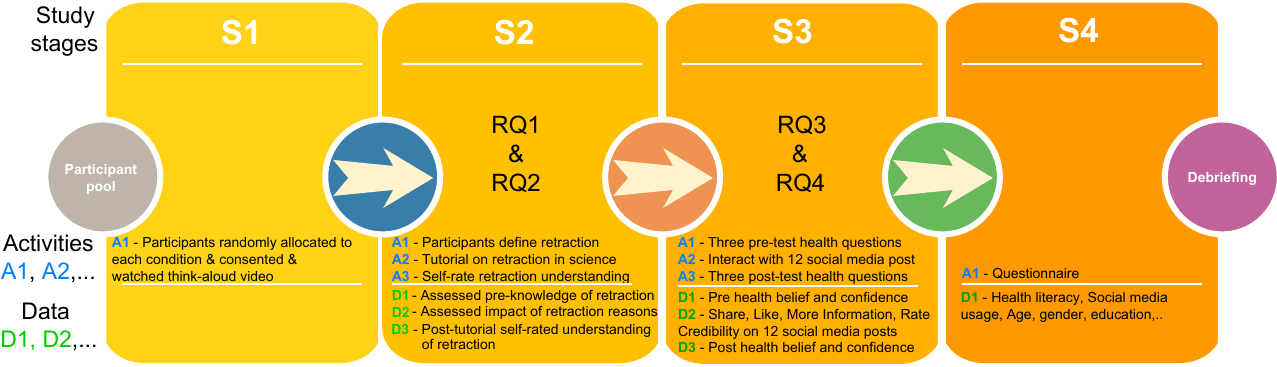}
    \caption{Overview of the study, its four stages are shown across the top of the figure. For each of them, we show the activities that participants worked through \textit{A1}, \textit{A2},.. and associated data we collected \textit{D1}, \textit{D2}. }
    \label{fig:study_flowchart}
\end{figure*}

Figure~\ref{fig:study_flowchart} provides an overview of the four stages in our study.
For each stage, the figure lists the activities participants did and the data collected. 
S1 is the preliminary set up stage.
Stage S2 activities are for the first two research questions on participants' understanding of retraction.
Stage S3 has the activities to address Research Questions 3 and 4
about changes in participants' beliefs after reading social media news posts.
In S2 and S3, we asked participants to ``think-aloud'' through each activity, to gain insights into their mental model, intentions and their understanding of the interface~\cite{gibson1997talking, nielsen2012thinking}.
The final stage, S4, had a questionnaire.
\color{black} The Treatment condition was different from Control only on SMNP interface $-$ their interface was augmented to give them access to retraction information about the published study. The rest of the activities were identical in both conditions.\color{black}


\subsection{Recruitment}

We recruited University of Anon first- and second-year psychology students who need to participate in research studies for course credit. 
We posted our study on the institutional recruitment platform, stating that participants must be aged 18 or more and fluent in English. The study was advertised as an online one-hour zoom session. 
\color{black}We recruited these students as they were the sample available to us. They had not learnt about retraction.
This population represents a relatively knowledgeable group, which will identify the minimum amount of support any population will require.
It also ensures variability in their prior experience to understand how experience matters.
Critically, this ensures against floor effects where no one in the entire sample has any knowledge, which is less informative for how knowledge may interact with scaffolding. \color{black}Furthermore, these participants have received the sort of training or education that high schools can deliver. \color{black}


\color{black}

\subsection{Stage S1: Preliminaries}
Upon arrival, each participant was automatically (randomly) assigned to either the Control or Treatment group. 
Then, they watched a video that demonstrated how to think aloud when using the interface in stage \textbf{S3}.

\subsection{Stage S2: Understanding of retraction - RQ1 and RQ2}

As shown in Figure~\ref{fig:study_flowchart}, this stage had three activities.
Activity \textit{A1} assessed participants' initial understanding of retraction (RQ1). 
\color{black} 
Then Activity \textit{A2} served two purposes:
(1) it was a tutorial about seven key reasons for retraction;
(2) as part of the active learning in the tutorial, participants rated the impact of each reason and the think-aloud provided insights about their reasoning.
The tutorial gave participants a common base knowledge of retraction which was needed for the rest of the study. \color{black}
We now explain the design of the activities \textit{A1} $-$ \textit{A3}.

In activity \textit{A1}, we first asked participants: \textit{How do you define the word retraction?} \color{black}This was an important baseline: not knowing the general meaning of the word is likely to have important implications for their knowledge of retraction in science publication context. \color{black}
We coded participants knowing it if they explained it broadly or with synonyms.
We then asked:\textit{What do you understand about retraction in a science publication context?} \color{black}This was important as the word retraction can be used in other contexts, for instance in journalism, where the meaning is different from scientific retraction.
\color{black}
Participants were rated as knowing this if they mentioned the withdrawal of a published paper.

\begin{figure}[ht]
    \centering
    \includegraphics[width=0.65\textwidth]{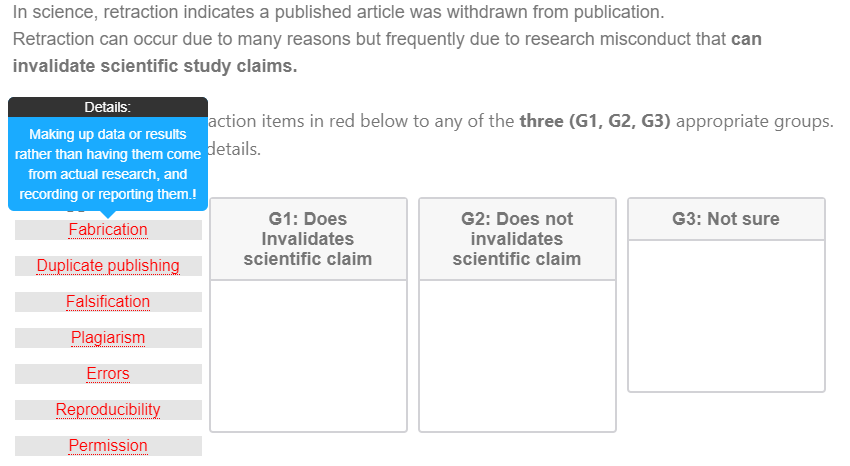}
    \caption{Activity \textit{A2} at start of tutorial. Participants  move each retraction reason (red underlined text) to a box, G1, G2 or G3, indicating if that reason invalidates the papers' scientific claims. As the user's mouse is over \textit{Fabrication}, a blue pop-up explains it.}
    \label{fig:tutorial_reasons_retraction}
\end{figure}

The \textit{A2} retraction tutorial activity is shown in Figure~\ref{fig:tutorial_reasons_retraction}.
The first sentence introduces the tutorial with a brief description of retraction in science publication.
\color{black} The second sentence states that retraction can occur for many reasons, including ones that can invalidate the scientific claims.
\color{black}
The tutorial asks the participant to move each retraction reason to one of the boxes at the right to show if that reason invalidates the paper claims.
As participants did this, they were asked to think aloud.
As in the screenshot, mousing over a reason presents a short explanation. A complete list of explanations and tutorial text is in supplementary Section~\ref{sec:appex_retract_tutorial}.
This activity served as a tutorial because participants actively considered each of these seven retraction reasons to decide whether it affects their trust in the scientific claims of a paper. 
\color{black}

In the final activity, A3, participants rated their understanding of retraction and its impact on the validity of a scientific claim on a scale of 1 to 10.

\subsection{Stage S3: Rating the credibility of news posts - RQ3 and RQ4}
Figure~\ref{fig:study_flowchart} show the three main activities in this stage.
RQ3 asks how a person’s prior belief about health topics affect their rating of the credibility of SMNP, that has information about retraction.
For this, we carefully chose health claims about three topics.
In Activity A1, participants answered questions indicating their agreement with each claim.
In Activity A2, they rated the credibility of 12 news posts, four per topic.
RQ4 asks how reading SMNP with retraction information changes people’s beliefs on health topics?
To assess this change, in Activity A3, participants again answered the questions from A1.  

As an introduction to this stage of the study, participants were informed that they would read and rate SMNPs.
They then did a familiarisation task with the interface used for Activity 2.
This had a post on curing brain cancer, a different topic from those in Activity 2.

In Activity \textit{A1}, participants rated their beliefs on the three health topics. First, they answered: \textit{How much do you agree with the following statements?} (with responses on a five point scale from Strongly Disagree to Strongly Agree). 

\begin{itemize}
    \item  \textit{Masks\footnote{in the rest of the paper these health topic claims will be referred to as italicised \textit{Masks, Diet, Movie}}:} Masks are effective in limiting the spread of coronavirus.
    \item  \textit{Diet:} Mediterranean diet is effective in reducing heart disease.
    \item  \textit{Movie:} Snacking while watching an action movie leads to overeating.
\end{itemize}

The second question asked \textit{How confident are you about your knowledge on the topic?}, with responses on a scale of 0 to 100. 
We chose these topics based on two key criteria.
First, we wanted readily understood topics on health and wellness.
Secondly, we chose topics so that participants would be likely to have different prior beliefs about each of them; this is needed to study the effects of prior belief (RQ3 \& RQ4).

To introduce Activity \textit{A2},
we  asked participants to 
\textit{Imagine yourself browsing through the posts as if they are showing up on your actual social media news feed.}.
It is important to encourage participants to use the interface as they would normally in~\cite{pennycook2020practical, yaqub2020effects}.
Each participant saw the same 12 SMNPs, 4 from each health topic but presented in random order. \color{black}The number headlines is in line with previous work on rating SMNPs~\cite{yaqub2020effects}\color{black}. 
To create the  12 SMNPs, we followed previous research guidelines which recommended using 
actual news headlines rather than artificial ones~\cite{pennycook2020practical}. \color{black} Following the guidelines, we aimed to collect real news SMNPs, each with the following attributes: 
(1) based on published studies; 
(2) with a clear scientific claim;
(3) with the base articles available online; and 
(4) having a distinct image.  
We were limited by what we found in the Retractionwatch database and Altmetrics database. Retractionwatch keeps records of retracted articles and Altmetrics tracks headlines disseminated based on those published and retracted articles.
We searched these databases for MMR vaccine, GMO, masks, diet, movie and overeating, gut bacteria, and neck injury due to texting. 
These were identified as health and wellness topics that the broad public can understand. 
We could only find SMNPs that met our criteria for three topics: \textit{Masks}, \textit{Diet} and \textit{Movie}. \color{black}
We initially aimed to find SMNPs both for and against each health claim and based on published papers, one retracted and one not.
However, we could only do this for the \textit{Masks} claim;
this is unsurprising as not all news becomes viral on social media~\cite{vosoughi2018spread}.
So the SMNPs we used were based on headlines disseminated from papers:
\begin{itemize}
    \item  \textit{Masks}: supporting ~\cite{zhang2020identifying} and refuting retracted\color{red} ~\cite{bae2020effectiveness}\color{black} ;
    \item  \textit{Diet}: supporting retracted~\color{red}~\cite{estruch2013primary}\color{black};
    \item  \textit{Movie}: supporting retracted \color{red}~\cite{tal2014watch}\color{black}, . 
\end{itemize}

\begin{figure}[H]

\centering
    
\begin{subfigure}{1\textwidth}
  \centering
  \includegraphics[width=0.86\textwidth]{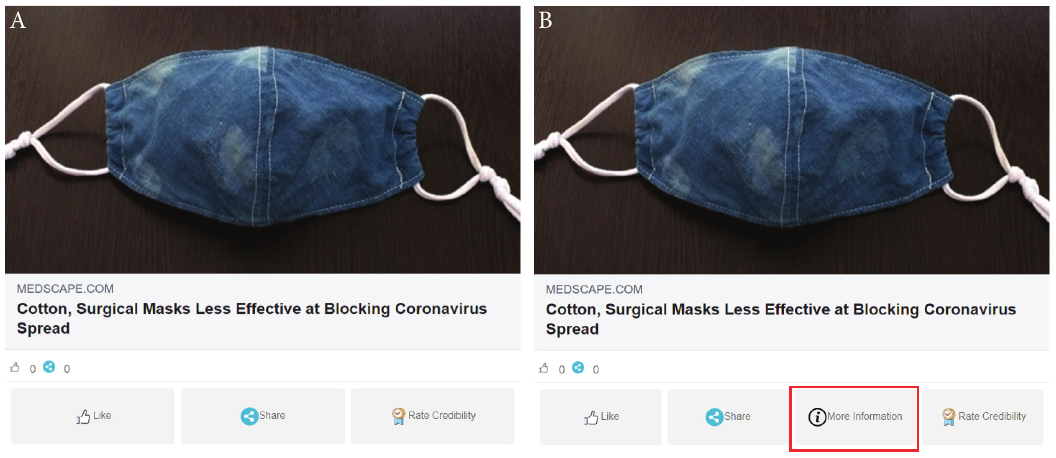}
  \label{fig:two_social_media}
\end{subfigure}
%

\vspace{-0.5cm}
\begin{subfigure}{0.42\textwidth}
  \centering
  \includegraphics[width=1\linewidth]{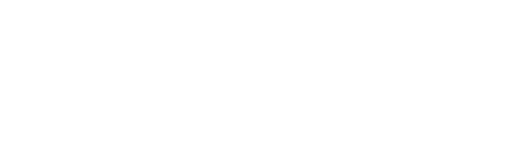}  
  \label{fig:retraction_info_revealed}
  
\end{subfigure}
\begin{subfigure}{0.42\textwidth}
  \centering
  \includegraphics[width=.90\linewidth]{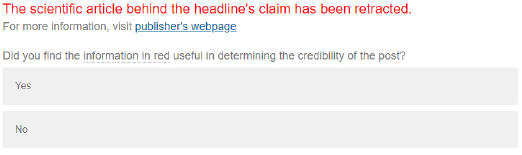}  
  \label{fig:retraction_info_revealed}
  
\end{subfigure}

\caption{\textbf{Stage S3 interfaces:}  (A) Control (B) Treatment. Both have buttons for \textit{Like}, \textit{Share}, and \textit{Rate Credibility}. (B) has also has \textit{More Information} (here with a red box) $-$ clicking this, brings up the information about retraction.  }
\label{fig:two_interfaces}
\end{figure}

Figure~\ref{fig:two_interfaces} shows the interfaces participants used to read the 12 SMNPs.
The Control group saw Interface A, on the left and
the Treatment group saw Interface B. 
\color{black}As per the widely used research guidelines~\cite{pennycook2020practical}\color{black}, these were designed to be similar to Facebook's news interface - this provides
an image, news source, headline and interaction buttons, 
\textit{Like} and \textit{Share}.
Critical to answer our RQs, we introduced the \textit{Rate Credibility} button. 
The only difference between A and B is the extra button for \textit{More Information}. (This figure adds a red box to highlight it). 
We placed this close to the other buttons because literature showed that people often do not notice the current Facebook \textit{More information} button~\cite{geeng2020fake}.

Both interfaces respond to clicks. 
Clicking anywhere on the post opens the underlying news article web-page in a new tab. Clicking on \textit{Like} and \textit{Share} updates the corresponding count (in the figure, they are 0). 
Clicking on \textit{Rate credibility} will ask participants to \textit{Please rate the credibility of the above social media post headline claim.} on scale of 0 to 10. 
A click on \textit{More information} reveals additional information shown in the figure.
The retraction status of the article  is in red font,
followed by with the link to the article on the publisher's site.
This is followed by a question about the usefulness of the red retraction information.



To assess any change in belief after reading the 12 SMNPs, participants in activity \textit{A3}, participants once again rated their beliefs on the three health topics (RQ4). 

\subsection{Stage S4: Final Questionnaire}


In this final stage (S4 in Figure~\ref{fig:study_flowchart}), participants completed a questionnaire about 
demographics, health literacy and social media use for news. \color{black}We included a question on health literacy since literature showed that people with lower literacy may be more prone to believe misinformation~\cite{damian2020promoting}, notably COVID-19 misinformation~\cite{bin2021covid, wojtowicz2020addressing}.\color{black}
This timing avoided any impact on the participants' responses in the earlier stages.

\color{black}
\subsection{Data Analysis}
During the study, we captured several forms of data, as summarised at the bottom of the boxes for Stages S2 $-$ S4 in Figure~\ref{fig:study_flowchart}.
For RQ1 and RQ2, in Stage S2 data, participants' responses were independently coded by two authors after the codebook had been created by the authors. Any disagreements were resolved through discussion.
For RQ3 and RQ4, the interface collected interface action as described above.
In all stages, participants were asked to think aloud.
Video of all sessions was recorded.
These recordings were transcribed and analyzed along with hand-written notes about observations by the first author who conducted the studies. 
We then followed an iterative coding process to analyze the think-aloud data. 
This was used to identify reasons for participants' interpretations of the interface and information they saw and for their explanations for their interpretations and actions.
For quantitative data in RQ3 and RQ4, we analyzed the data using Linear Mixed Effects Regression (LMER). Our outcome variable is the credibility ratings of the SMNP based on different factors. The dependent variable is on a scale of 0 to 10, indicating the SMNPs credibility ratings. The LMER model included random effects for participant ID and headline ID, in order to account for the repeated measures of different items. In addition to demographic factors such as 
age and gender, 
the independent variables included 
the group (treatment/control), health topic, 
political affiliation, 
and social media usage.


We used the stepwise approach described by Winter~\cite{winter2013linear} and others~\cite{sscc-mmt} to incorporate the 
various independent variables with corresponding interaction effects. We followed standard practices for model selection and significance testing for mixed models by using the Likelihood Ratio Test (LRT) which tests the difference between two nested models using the Chi-square test. First, we compared the initial constant model (with no predictors) with the model that includes predictors. We retained a predictor as long as it exhibited statistical significance compared to the initial model. In each step, we included the participant ID and headline direction as random effects to account for repeated measures for varying items~\cite{winter2013linear,brauer2018linear}. We further conducted posthoc analysis for different subgroup comparisons.

\color{black}


\section{Results}

\label{sec:results}

This section is organised around the four research questions.
For each of these, we report both quantitative data along with qualitative think-aloud information.
Throughout these sections, we refer to participants as in this example:
\color{black}\textbf{P9$_{C_{00}}$}\color{black}.
This refers to Participant 9. 
The first subscript can be C or T, for the Control or Treatment condition. 
The next pair of subscripts show whether the participant was able to explain the meaning of retraction in Stage S2-A1.
The values can be 0 or 1.
In this example, the first 0 means \textbf{P9} could not explain the broad meaning, and the second 0 means they could not explain it in the context of science publications.

\subsection*{Participants}
We recruited 44 participants.
Most (40, 91\%) were aged 18-25; the other four were 26-35. 
Thirty were female and 14 male, representative of psychologists in Australia~\cite{black2019relationship}. 
All lived in Sydney, Australia, with exception of \color{black}\textbf{P7$_{C_{00}}$}\color{black} in Hong Kong. 
All had completed high school, but 3 had also completed a degree. 
Seven participants (16\%) had low health literacy. 
Only one participant, \color{black}\textbf{P9$_{C_{00}}$}\color{black}, had no social media account. 
The rest had used social media for three or more years. 
Of the 43 social media users, 38 used it often, 4 sometimes and 1 hardly ever. 
For those who reporting using it often, their estimated daily use for news was 52 minutes.
Platforms used most frequently were 
Instagram (by 37 $-$ 84\% of participants),
Facebook (by 20 $-$ 45\%) and 
SnapChat (by 17 $-$ 39\%).
Overall, this indicates that all but one participant had long term and regular use of social media platforms with a similar interface to the one in the study.
The news sources participants trusted did not align with those most used.
Eighteen (41\%) trusted print newspapers but just two used them.
For television news, nine (20\%) trusted it and 22 (50\%) used it.
For social media, the corresponding numbers are two (5\%) for trust and 17 (39\%) for use.
There was consistency for weblogs with 26 (59\%) trusting and 34 (77\%) using them.
This paints a complex picture where many participants commonly consume news from multiple sources, including ones they do not consider trustworthy.

\subsection{RQ1: How do people understand the retraction of scientific publications?}
\label{sec:RQ1_retraction_understanding}

\begin{figure}[ht]
    \centering
    \includegraphics[width=1\textwidth]{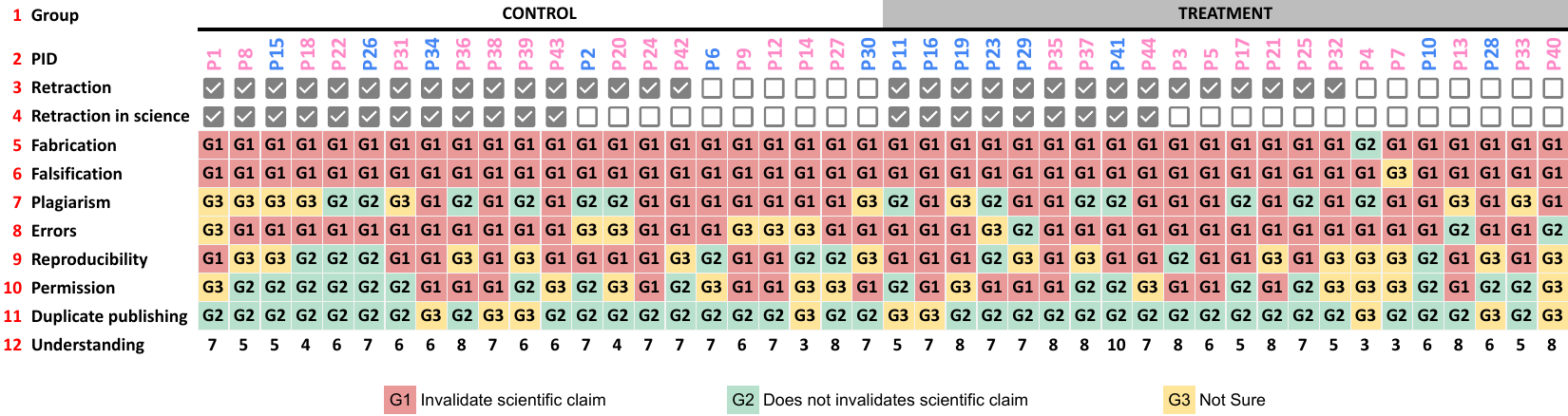}
    \caption{Participant understanding of retraction. Row 1 is the group, Control (left) or Treatment (right). Row 2 shows the Participant ID (male-blue, female-pink). Rows 3 and 4 show if participants could explain retraction in general and in the context of scientific publication. Rows 5-11 shows the reasons for paper retraction, and each cell indicates whether that participant indicates whether that reason invalidates the findings of the paper only Rows 5-7 are fraudulent. Row 12 is participant self-rating of understanding retraction at the end of the tutorial. Summative table of reasons for retractions is in the appendix}
    \label{fig:tutorial_retract_response}
\end{figure}

Rows 3 and 4 of Figure~\ref{fig:tutorial_retract_response} summarise the results.
The Activity \textit{A1} participant responses for Retraction and Retraction in science in Figure~\ref{fig:tutorial_retract_response} were double coded by two authors. 
Only two responses had disagreement, and were resolved through discussion. 
Row 3 of the figure shows that similar numbers of participants understood the general meaning of retraction in the Control (16) and Treatment (15) groups. 
Some examples of answers include:
\textit{``removal of something"}\footnote{All quotes have been revised to remove disfluencies and to add clarifications in square brackets.
}\textbf{(\color{black}P3$_{T_{10}}$\color{black})}, \textit{``taking something back already being said"}\textbf{(\color{black}P15$_{C_{11}}$\color{black})}

For the case of scientific retraction, Row 4 in Figure~\ref{fig:tutorial_retract_response} shows that about half the participants demonstrated understanding, with more in the Control (12) than Treatment (9) group. 
These 21 were a strict subset of the 31 who understood the term in general.
Examples of definitions we coded as correct include:\vspace{0.2cm}

\textit{``fake science that got through the cracks and got published and it would be pulled back"}\textbf{(\color{black}P1$_{C_{11}}$\color{black})},
\textit{``when something is retracted, I guess it is found to be not credible and so the authors retracted their scientific paper, statements or published research"}\textbf{(\color{black}P41$_{T_{11}}$\color{black})}.
\vspace{0.2cm}

Notably, most participants who correctly explained retraction in science mentioned that it was due to flaws in the method or analysis. Essentially, they described retraction reasons that can be categorised as fabrication, falsification, reproducibility and error. None mentioned duplicate publishing, plagiarism or permission as a reason for retraction. 



Overall, this stage indicated that most participants (31 $-$ 70\%) could define retraction in general.
About half (21 $-$ 48\%) could do this for a scientific context and they
most often referred to flaws in the study that caused its retraction.
The Control and Treatment groups had similar levels of understanding,
although the Control group had somewhat more participants showing understanding of science retraction.

\subsection{RQ2: What reasons for retractions do people consider to impact credibility of scientific claims?}
\label{sec:RQ2_reasons_retraction}

Rows 5 $-$ 11 in
Figure~\ref{fig:tutorial_retract_response} show data from activity \textit{A2} in Stage \textbf{S2} in Figure~\ref{fig:study_flowchart}.
The coding shows: 
\colorbox{red!15}{G1} where the participant indicates that this reason  invalidates the paper claim;  
\colorbox{green!20}{G2} when it does not;
and \colorbox{yellow!25}{G3} is for unsure.

\subsection*{Fraudulent reasons for retraction}
The rows for \textbf{\textit{Fabrication}} and \textbf{\textit{Falsification}}
are dominated by \colorbox{red!15}{G1} cells. 
Participants comments referred to the definitions in tutorial hover text.
Three participants referred to the Karl Popper falsification principle as in this comment:
\textit{``reminds me of falsification theory"} \textbf{(\color{black}P3$_{T_{10}}$\color{black})}.

The last fraudulent retraction reason, \textbf{\textit{Plagiarism},}
had a far more mixed response:
\colorbox{red!15}{G1}=22, \colorbox{green!20}{G2}=13, and \colorbox{yellow!25}{G3}=9.
Comments explaining the \colorbox{red!15}{G1} (invalidates) response linked it to other problems as in this example:
\vspace{0.2cm}

\textit{``I mean if you are already using someone else`s idea and not even referencing them already, pretty much shows that you thought you found these results on your own, those results could have been manipulated."} (\color{black}\textbf{P14$_{C_{00}}$}\color{black}).\vspace{0.3cm} 

Comments from the thirteen with \colorbox{green!20}{G2} responses argued that a scientific study could still be valid when the researcher has plagiarised it from any other valid and reputable study, as in this example: \vspace{0.2cm}

\textit{``Plagiarism, I guess if it is still coming from [valid study], this would be more like [an] incorrect reference rather than invalidating its own claim. As long as the data you use or idea is correct"} \textbf{(\color{black}\textbf{P11$_{T_{11}}$}\color{black})} 
\vspace{0.2cm}

Similar views were made by those who were unsure (\colorbox{yellow!25}{G3}): \vspace{0.2cm}

\textit{``not sure if plagiarism would [invalidate scientific claim], I am not sure, I mean plagiarism obviously is bad, but does that necessarily invalidates scientific claim, I mean you could plagiarise something but still could be true, right, so I am not sure. That paper could be retracted because someone stole the idea, but it does not necessarily invalidate the claim"}.\vspace{0.2cm}

None of the 44 participants mentioned the possibility of unintentional or self plagiarism.

\subsection*{Non-Fraudulent reasons for retraction}
Responses for these are in rows 8 $-$ 11 of Figure~\ref{fig:tutorial_retract_response}.
Row 8, for \textbf{\textit{Errors}}, shows that most
participants (34 $-$ 77\%) rated this \colorbox{red!15}{G1}.
Of the 34, 14 justified this in terms of questioning the method and findings of the study and considered errors in research to invalidate scientific claims. 
Three (3) participants considered errors did not invalidate the paper results (\colorbox{green!20}{G2}) and 7 were unsure.
Their comments referred to the nuances in errors and 
that no research is completely without errors such as systematic, random, or human errors, as in this example: \vspace{0.2cm}

\textit{``I think errors in research are bound to happen, as much as you try to control, something or the other is bound to go wrong"}(\color{black}\textbf{P13$_{T_{00}}$}\color{black}). \vspace{0.2cm}

Row 9 shows the diverse ratings for \textbf{\textit{Reproducibility}}: 
\colorbox{red!15}{G1} $-$ 21;
\colorbox{green!20}{G2} $-$ 14;
\colorbox{yellow!25}{G3}  $-$ 9.
Our participants had all been taught about the reproducibility crisis~\cite{ioannidis2005most}
in psychology.
Seven mentioned this at this point $-$
for example, \color{black}\textbf{P16$_{T_{11}}$}\color{black} said: \textit{``I would say yes [it invalidates scientific claim] as well, I know there is a big thing talked about in social psychology in particular in reproducibility"}.
Surprisingly, these participants
had a spread of ratings of its impact on credibility,
three with \colorbox{red!15}{G1} ratings,
(\color{black}\textbf{P16$_{T_{11}}$}\color{black}, 
\color{black}\textbf{P25$_{T_{10}}$}\color{black}, 
 \color{black}\textbf{P21$_{T_{10}}$}\color{black}), 
two with \colorbox{green!20}{G2},
(\color{black}\textbf{P18$_{C_{11}}$}\color{black}, 
 \color{black}\textbf{P22$_{C_{11}}$}\color{black}) 
and two with \colorbox{yellow!25}{G3} 
(\color{black}\textbf{P28$_{T_{00}}$}\color{black}, and 
\color{black}\textbf{P29$_{T_{11}}$}\color{black}). 
Five of those who were uncertain said they were unsure of the meaning and ten participants pronounced it Reproducibility (with a \textit{k} sound instead of the \textit{s} sound).
This suggests they were not very familiar with the word.
Just one of them rated it as \colorbox{red!15}{G1}
(\color{black}\textbf{P3$_{T_{10}}$}\color{black}). 
The rest were nearly evenly split on their ratings, with four of \colorbox{green!20}{G2}
\color{black}\textbf{P26$_{C_{11}}$}\color{black}, 
\color{black}\textbf{P33$_{T_{00}}$}\color{black}, 
\color{black}\textbf{P39$_{C_{11}}$}\color{black}, 
\color{black}\textbf{P42$_{C_{10}}$}\color{black}  
and 5 of \colorbox{yellow!25}{G3},
\color{black}\textbf{P4$_{T_{00}}$}\color{black}, 
\color{black}\textbf{P6$_{C_{00}}$}\color{black}, 
\color{black}\textbf{P14$_{C_{00}}$}\color{black},  
\color{black}\textbf{P32$_{T_{10}}$}\color{black}, 
\color{black}\textbf{P40$_{T_{00}}$}\color{black}. 
Half of these participants could not define the general meaning of the retraction and 8 could not define it in the scientific context.


Responses about \textbf{\textit{Permission}} (Row 10 in Figure~\ref{fig:tutorial_retract_response}) were spread (\colorbox{red!15}{G1}=15, \colorbox{green!20}{G2}=17, and \colorbox{yellow!25}{G3}=12).
Four out of 15 participants who rated it as \colorbox{red!15}{G1},
cited ethical reasons, for examples:
\textit{``Yeah, I am just gonna say that [permission retraction reason] invalidates scientific claims because it's unethical."} (\color{black}\textbf{P29$_{T_{11}}$}\color{black}) 
The 17 participants responding \colorbox{green!20}{G2} commented on validity of the underlying data as in this example: 
\textit{``[\colorbox{green!20}{G2}] because the data may still be correct."} (\color{black}\textbf{P2$_{C_{10}}$}\color{black}) \color{black}
Similar views were echoed by 12 participants giving the rating \colorbox{yellow!25}{G3}, for example:
\textit{``I don't know if it invalidates the scientific claims or not necessarily because the data could still be true."} (\color{black}\textbf{P44$_{T_{11}}$}\color{black}).
Participants who gave all three \colorbox{red!15}{G1} ratings spoke about unethical data collection and that the actual data and claims may be valid.

For \textbf{\textit{Duplicate publishing}} (Row 11), 
most participants (35 out of 44, $\approx$ 80\%) rated it \colorbox{green!20}{G2}, with the other 9 unsure (\colorbox{yellow!25}{G3}).
Five of them related it to plagiarism or misconduct.
Nine made comments that it indicates double validation of the article,
for example:\vspace{0.3cm}

\textit{``I think [duplicate publishing] does not invalidate a scientific claim because I am assuming if that publication has gone through the [review process] twice [and] through the rigorous vetting that scientific data [and a] scientific report goes through, it means that it has gone through the correct channels."}
(\color{black}\textbf{P13$_{T_{00}}$}\color{black}) and  
\textit{``that does not invalidate that would validate your findings because it can be accepted multiple different places"}(\color{black}\textbf{P43$_{C_{11}}$}\color{black}).

\subsection*{Participant self-rated knowledge,  Stage S2 - A3}

Having completed the tutorial, 
participants' self-rated their
understanding of retraction and its impact on the validity of a scientific claim.
The last row in Figure~\ref{fig:tutorial_retract_response} 
show these.
The mean was 6.4 (median and mode 7) out of the 10 maximum rating.
Most participants explained their score in terms of what they learnt in the tutorial, for example:\vspace{0.3cm}

\textit{``From the above [tutorial], it is quite clear why you would retract a paper for misconduct. [Now] I am gonna say an 8, I feel like there might be things that I didn't understand before [the] tutorial [and] I would have given myself 2 before [the] tutorial."}(\color{black}\textbf{P3$_{T_{10}}$}\color{black} score 8).

Three participants had ratings of 3:
\color{black}\textbf{P4$_{T_{00}}$}\color{black},
with the only non-\colorbox{red!15}{G1} rating for Fabrication,
\color{black}\textbf{P7$_{T_{00}}$}\color{black}, 
the only non-\colorbox{red!15}{G1} rating for Falsification and
\color{black}\textbf{P14$_{C_{00}}$} \color{black} who said:
\textit{``I feel like a lot of information was given about retraction and I do not fully understand what it means. You can see that I was guessing a lot of it. [I] did not have enough reasoning to say why I put them in because I am not fully understanding them."}

Overall, the tutorial meant that most participants moved to the next phase with awareness and some understanding of retraction and its impact on credibility of claims in published claims.

\color{black}
\subsection*{Summary}

Table~\ref{tab:summative_table_retract} summarises the results for RQ1 and RQ2,
aggregating the detailed picture in Figure \ref{fig:tutorial_retract_response}.
This shows participant retraction understanding and how participants classified each reason for retraction.
The first two rows summarise the understanding that participants could demonstrate when asked to explain the term retraction in general and then in the context of scientific publications.
Around 70\% of participants could define the term in general.
For the scientific context, 55\% of the control group could explain it and 41\% of the Treatment group.

The next block of tallies shows counts of participants rating for each reason for retraction.
Underlined counts indicate broad consensus.
Only the first three reasons are fraudulent.  
There was strong consensus that 
Fabrication, Falsification and Errors do impact the validity (\textbf{\colorbox{red!25}{G3}}) and 
that Duplicate Publishing does not (\textbf{\colorbox{green!20}{G2}}). 
Other reasons have much more spread in participant ratings, although around half of the participants saw plagiarism and reproducibility as compromising the validity of claims, with about a quarter more of the participants being uncertain.
The last part of the table shows the average of the participants' self-rating of understanding of retraction at the end of the tutorial,
on a scale of 1 to 10.

\color{black}


\begin{table}[H]
\centering
\caption{\color{black}Summary of RQ1 and RQ2 results. (A) Number of participants who understood retraction, in general and in scientific publication. (B) Counts of participants who assessed each retraction reason invalidates paper findings. (C) Self-rated understanding of retraction after the tutorial.\color{black}}
\label{tab:summative_table_retract}
\begin{tabular}{lccccclc}
\multicolumn{8}{c}{\cellcolor[HTML]{EFEFEF}\textbf{(A) Understanding - RQ1}}                                                                                                                             \\
                         & \multicolumn{3}{c}{\textbf{Control (N=22)}}                        & \textbf{}            & \multicolumn{3}{c}{\textbf{Treatment (N=22)}}                                 \\
Retraction               &                      & 16                   &                      &                      &                      & \multicolumn{1}{c}{15} & \multicolumn{1}{l}{}          \\
Retraction in Science    &                      & 12                   &                      &                      &                      & \multicolumn{1}{c}{9}  & \multicolumn{1}{l}{}          \\
\multicolumn{8}{c}{\cellcolor[HTML]{EFEFEF}\textbf{(B) Tutorial - RQ2}}                                                                                                                                  \\
\textbf{Reasons:}        & \textbf{\colorbox{red!15}{G1}}          &                      & \textbf{\colorbox{green!20}{G2}}          & \textbf{}            & \textbf{\colorbox{yellow!25}{G3}}          &                        & \multicolumn{1}{l}{\textbf{}} \\ \cline{2-2} \cline{4-4} \cline{6-6}
Fabrication              & \underline{\textbf{43}}                   &                      & 1                    &                      & 0                    &                        &                               \\
Falsification            & \underline{\textbf{43}}                   &                      & 0                    &                      & 1                    &                        &                               \\
Plagiarism               & 22                   &                      & 13                   &                      & 9                    &                        &                               \\
Errors                   & \underline{\textbf{34}}                   &                      & 3                    &                      & 7                    &                        &                               \\
Reproducibility          & 21                   &                      & 9                    &                      & 14                   &                        &                               \\
Permission               & 15                   &                      & 17                   &                      & 12                   &                        &                               \\
Duplicate publishing     & 0                    &                      & \underline{\textbf{35}}                   &                      & 9                    &                        &                               \\
\multicolumn{8}{c}{\cellcolor[HTML]{EFEFEF}\textbf{(C) After Tutorial}}                                                                                                                                  \\
Self-rated understanding (max=10) & \multicolumn{3}{c}{6.2}                                            &                      & \multicolumn{3}{c}{6.6}                                                       \\
\textbf{}                & \multicolumn{1}{r}{} & \multicolumn{1}{l}{} & \multicolumn{1}{r}{} & \multicolumn{1}{l}{} & \multicolumn{1}{r}{} &                        & \multicolumn{1}{l}{}         
\end{tabular}
\end{table}







\subsection{RQ3: How does a person’s prior beliefs about a topic affect their rating of the credibility of SMNP that has information about retraction?}

This section reports results from the two think-aloud activities in Stage S3.
From Activity \textit{A1}, we report participants' prior beliefs in each of the three health claims,
highlighting the differences.
From \textit{A2}, where participants rated the credibility of 12 SMNPs, we report the differences between Control and Treatment groups for each of the topics. 
Then we summarise the level of use of \textit{Like}, \textit{Share} and \textit{More Information}.

\subsection*{Initial beliefs about the topics and confidence}

\begin{figure}[h]
     \centering
     \begin{subfigure}[b]{0.27\textwidth}
         \centering
         \includegraphics[width=\textwidth]{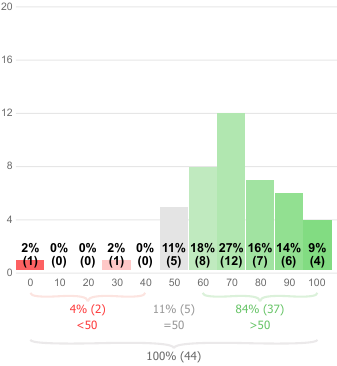}
         \caption{Pre \textit{Masks}}
         \label{fig:pre_mask_conf}
     \end{subfigure}
     \hfill
     \begin{subfigure}[b]{0.27\textwidth}
         \centering
         \includegraphics[width=\textwidth]{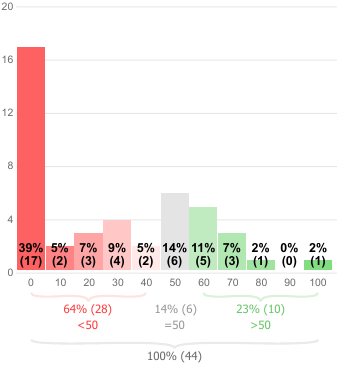}
         \caption{Pre \textit{Diet}}
         \label{fig:pre_diet_conf}
     \end{subfigure}
     \hfill
     \begin{subfigure}[b]{0.27\textwidth}
         \centering
         \includegraphics[width=\textwidth]{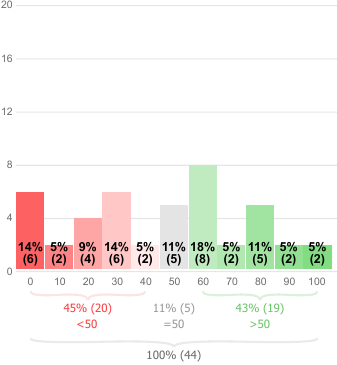}
         \caption{Pre \textit{Movie}}
         \label{fig:pre_movie_conf}
     \end{subfigure}
     
             \caption{Confidence distributions for the ratings in Table \ref{tab:pre_stance}}
        \label{fig:pre_three_health_confidence}
\end{figure}

Table~\ref{tab:pre_stance} shows the response distribution of participants' prior beliefs for the three topics, \textit{Masks, Diet, and Movie}. Figure~\ref{fig:pre_three_health_confidence} shows the confidence distributions. 
In line with our study design, participants prior beliefs are different across these topics.\newline

\begin{table}[h]
\caption{Prior belief ratings on: \textit{Masks:} Masks are effective in limiting the spread of coronavirus.; \textit{Diet:} Mediterranean diet is effective in reducing heart disease.; \textit{Movie:} Snacking while watching an action movie leads to overeating. The mean and median are for a scale from -2 (Strongly Disagree) to +2 (Strongly Agree). Bold underline indicated the modal value in a topic row.} 
\label{tab:pre_stance}
\resizebox{0.7\columnwidth}{!}{
\begin{tabular}{lccccc|cc}
\\
 & \multicolumn{1}{c}{\textbf{\begin{tabular}[c]{@{}c@{}}Strongly \\ Disagree\end{tabular}}} & \multicolumn{1}{c}{\textbf{\begin{tabular}[c]{@{}c@{}}Somewhat\\ Disagree\end{tabular}}} & \multicolumn{1}{c}{\textbf{Neither}} & \multicolumn{1}{c}{\textbf{\begin{tabular}[c]{@{}c@{}}Somewhat \\ Agree\end{tabular}}} & \multicolumn{1}{c}{\textbf{\begin{tabular}[c]{@{}c@{}}Strongly\\ Agree\end{tabular}}} & \multicolumn{1}|{l}{\textbf{~Mean}} & \multicolumn{1}{l}{\textbf{Median}}  \\
\textbf{\textit{Masks}} & 0 & 0 & 1 & 7 & \underline{\textbf{36}} & 1.8 & 2  \\
\textbf{\textit{Diet}} & 0 & 3 & \underline{\textbf{25}} & 15 & 1 & 0.3 & 0  \\
\textbf{\textit{Movie}} & 3 & 5 & 8 & \underline{\textbf{25}} & 3 & 0.5 & 1 
\end{tabular}
}
\end{table}

\subsubsection*{\textbf{Masks.}}
Table~\ref{tab:pre_stance} shows that almost all participants (43 of 44) agreed that:
\textit{Masks are effective in limiting the spread of coronavirus}.
(\color{black}\textbf{P8$_{C_{11}}$}\color{black}), 
\color{black}\textbf{P31$_{C_{11}}$} \color{black} selected `Neither', 
with low confidence, commenting: 
\textit{``I already heard differing opinions, although I have heard differing opinions, I have not researched it myself".} 
Adding to this picture, Figure~\ref{fig:pre_three_health_confidence}a
shows participants' confidence in their answers.
Most were confident, indicated by the dominance of high (green) confidence scores
(mean = 70.5, SD = 19.3). 
Participants explained their rating, referring to information from authoritative bodies and news media.
Examples include: \vspace{0.3cm}

\textit{``Yep, I strongly agree [with] this, definitely scientific evidence backs this up, yes it's [masks] definitely effective."} (\color{black}\textbf{P3$_{T_{10}}$}\color{black}), 
\textit{``I would say strongly agree based off different regulations [and] stuff put in place by institutions like WHO. I guess I trust their scientific background and research output in [assessing] the effectiveness of masks."}

\subsubsection*{\textbf{Diet.}}
For this topic, Table~\ref{tab:pre_stance} shows that the dominant ratings are 
`Neither', from 25 participants (57\%)
and `Somewhat Agree' from 15 (34\%).
The mean, at 0.3 is closest to the `Neither' rating (SD 0.6).
Confidence scores in Figure~\ref{fig:pre_three_health_confidence}b
show a strikingly low confidence with 17 (39\%) giving a 0 score and 64\% with confidence scores below 50 $-$ mean = 28.9, SD = 28.8.
Together, these indicate that many participants did not know about this topic.
Only \color{black}\textbf{P12$_{C_{00}}$} \color{black} had a `Strongly Agree' rating and commented:\textit{``I think strongly [Agree] and I have read some journals and also have taken a nutrition class and it [the Mediterranean diet] was covered in [class] lecture's content."}
Of the 25 with a `Neither' rating, 21 (48\%) said they were unaware of the topic and the factors involved in the diet and heart disease. Examples include:\vspace{0.3cm}

\textit{``I don't know what the Mediterranean diet is [nor do I have] the knowledge of the topic."} (\color{black}\textbf{P23$_{T_{11}}$}\color{black}) and 
\textit{``I know that the Mediterranean diet is considered relatively healthy, but [I do] not necessarily know what parts of the diet are beneficial in comparison to other diets. Neither, I am going to say neither because I don’t know how effective it is in reducing heart disease."} (\color{black}\textbf{P26$_{C_{11}}$}\color{black}).\vspace{0.3cm}

\subsubsection*{\textbf{Movie.}}
The third row of Table~\ref{tab:pre_stance} shows 25 participants somewhat agreed (57\% )
and the mean rating of 0.5 (SD 1.0) is between this and `Neither'. 
The confidence in these ratings is spread across the scale in Figure~\ref{fig:pre_three_health_confidence}c (mean = 45.68, SD = 29.68).
%
%
There were 20 participants with confidence $<50$ and a similar number, 19 had scores $>60$.
%
Comments include: \vspace{0.3cm}

\textit{``Because that is the story of my life, not practising  mindful eating very easily, just keep going, and I think I am confident of the topic. It is my personal area of interest, and I am prone to it."} (\color{black}\textbf{P13$_{T_{00}}$}\color{black}),
and \textit{``agree, do it every time, my own anecdotal evidence, whenever I watch a movie I end up eating way too much."} (\color{black}\textbf{P43$_{C_{11}}$}\color{black}). \vspace{0.3cm}

\subsection*{Summary of prior beliefs}

Overall, this analysis indicates that the three topics have very different profiles of initial participant beliefs:
for \textit{Masks}, participants mainly agreed and were confident;
for \textit{Diet}, most they did not know enough to make a judgement;
and \textit{Movie} was between these two in that some people thought it seemed reasonable, but there was a broad spread of confidence ratings.
A one-way ANOVA test gave $p<0.001$.

\subsection*{Credibility ratings across topics}

We now report the results for activity A2 where participants rated 12 SMNPs.
523 ratings in all, 174 responses for \textit{Masks} and \textit{Movie}, individually, and 175 for Diet (as participants skipped 2, 1 and 2 respectively).
Figure~\ref{fig:cred_ratings} summarises the results, in pairs of distributions showing the responses of the control group compared with the Treatment group.
Figure~\ref{fig:cred_ratings}a does this for the \textit{Masks} topic,
separating results for the 2 posts that supported the claim that masks are effective and the 2 that were negative.
Figure~\ref{fig:cred_ratings} has the results for \textit{Diet}, and \textit{Movie} topics, where all 4 SMNPs supported the claims.
Left of each distribution visualisation, we report the mean, median and Standard Deviation.
This figure integrates the prior beliefs of the participants with their credibility rating of the SMNP on that topic. A point's y-axis value represents its credibility rating; a point's color represents prior belief. The figure shows the strong relationship between the two ratings.
The green dots are for participants who agreed, either somewhat (+1) or strongly (+2).
Red dots are for those somewhat (-1) or strongly (-2) disagreed and the grey dots are for ratings of neither (0).
The means for each of these subgroups is shown at the top of Figure~\ref{fig:cred_ratings}b. Critically, this figure also separates credibility ratings of the Control group, and the Treatment group, which had retraction information available. Here we can see the direct effect of retraction information on the ratings. \color{black} Step1 in the stepwise regression analysis shown in section~\ref{sec:stepwise_regression} indicates that the independent variable group had an impact in reducing the ratings, with $p-value<0.001$. The impact of retraction is also shown in the posthoc analysis using different health topic levels $p-value<0.01$. The p values were adjusted using Bonferroni method. This can be seen in appendix section ~\ref{sec:subgroup_compare}
It is worth noting that none of the non-retracted supporting masks SMNPs received ratings lower than 5 in Treatment. This observation is also clearly visible from the distribution plot in Figure~\ref{fig:cred_ratings} as none of the data points under \textit{Masks} positive(T) plot goes below 5.
\color{black}

\subsubsection*{\textbf{Masks.}}
\begin{figure}[]
\begin{subfigure}{0.8\textwidth}
  \centering
  \caption{}
  \includegraphics[width=0.9\linewidth]{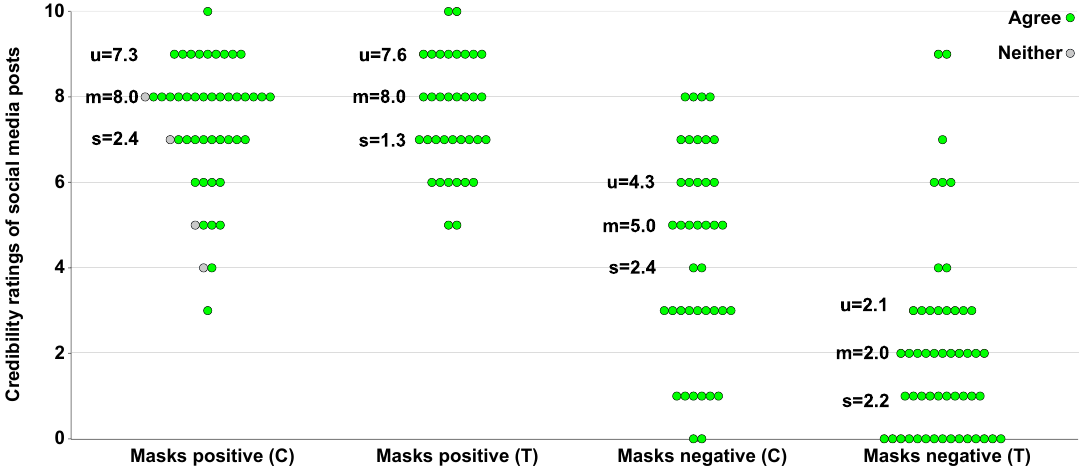}
  \label{fig:mask_cred}
\end{subfigure}
%


\begin{subfigure}{0.8\textwidth}
  \centering
  \caption{ }
  \includegraphics[width=0.9\linewidth]{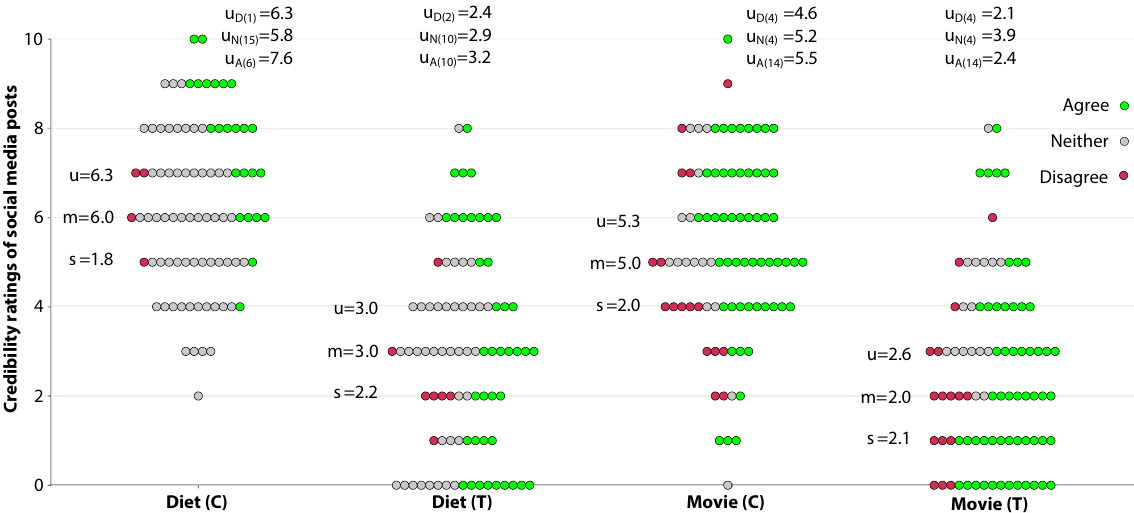}  
  \label{fig:diet_movie_cred}
\end{subfigure}

\caption{Participants credibility ratings for (a) Masks SMNPs - positive posts left, negative right (b) Diet and Movie. Each has distributions for Control (C) and Treatment (T). Each dot shows one rating. The colour shows the prior belief: red $-$ disagree; grey  $-$ neither; and green $-$ agree. Left of each distribution is its mean (u), median (m) and standard deviation (s). Above each distribution in (b) are details according to prior beliefs $-$ for example, over Diet (C), $u_{N(15)}=5.8$ is the mean for the 15 participants with prior belief rating of \textit{Neither}.  }
\label{fig:cred_ratings}
\end{figure}

In all four distributions in Figure~\ref{fig:cred_ratings}(a), we see the dominance of participants' prior belief that masks are effective, with almost all the dots being green.
The left pair of distributions has the results for the 2 SMNPs that supported the claim and they were based on publications that were \textit{not} retracted.
Control and Treatment groups have very similar credibility rating distributions, and both have a high credibility, with means of 7.3 and 7.6 respectively.

The pair of distributions at the right show the impact of making retraction information available for the control group. 
Here, the 2 SMNPs claiming that masks are ineffective both have very low credibility ratings.
For the control group the mean is 4.3.
By contrast, in the Treatment group, the mean is just 2.1.
There is also a striking number of participants who gave a rating of zero. \color{black}Additionally, there are two  SMNP evaluations by \color{black}\textbf{P40$_{T_{00}}$}\color{black}\color{black} that were  rated as 9 on the credibility scale in the Masks Negative (T) Treatment condition. For both of these data points the retraction information was not accessed by \color{black}\textbf{P40$_{T_{00}}$}\color{black}.

\subsubsection*{\textbf{Diet.}}

In this topic, where most participants had little prior knowledge, 
Figure~\ref{fig:cred_ratings}b shows the strong impact of access to retraction information in the Treatment group.
Overall, the control group credibility ratings, Diet (C), the mean is 6.3, double that for Diet (T) at 3.0.
The figure shows a corresponding difference for each prior-belief group,
both in the distributions of each colour and the means at the top of the figure.
For example, in Diet (C), the 15 participants had a prior belief rating of \textit{Neither} (grey dots) and their mean rating for the SMNPs was 5.8, double that for the 10 Diet (T) participants with prior belief of \textit{Neither} ($u_{N(10)}=2.9$).

\subsubsection*{\textbf{Movie.}}

The right pair of distributions in Figure~\ref{fig:cred_ratings}b, for the Movies topic
is dominated by the 28 green dots of participants with prior agreement.
Here, the control group has a mean SMNP credibility rating of 5.3 where
the Treatment group mean is 2.6. 
There is a drop across each three prior belief sub-group;
for example, for the 28 people with prior agreement,
the 14 in the Control group had a mean credibility rating of 5.5
where the Treatment group mean was only 2.4.

\subsection*{Credibility Ratings Summary.}
The results show a consistent and meaningful difference between the Control and Treatment groups
for the 2 SMNPs negating the \textit{Masks} claim and for the 4 SMNPs for both \textit{Diet} and \textit{Movie}.
Because the Treatment group had access to information about retraction, they were strongly influenced to downgrade the credibility of these SMNPs.
The effect of Treatment group to reduce credibility was consistent across the distribution of prior beliefs.

\subsection*{Retraction Knowledge interaction with credibility ratings.}
We also analysed the interaction between participant retraction knowledge, measured in RQ1, and how much participants downgraded the credibility of SMNPs when provided with retraction information. It is important to evaluate how the effect of the Treatment depends on prior knowledge because it suggests how the degree to which further education about retraction is needed for retraction information on social media to be useful to the broader population.
Table \ref{tab:my-table} shows the SMNP ratings separated into three groups based on their retraction knowledge. Retraction knowledge is broken into two kinds: general knowledge and knowledge in science. Having adequate knowledge is represented by a 1, and a lack of knowledge is 0. The first column shows the mean credibility ratings for participants who could not explain retraction in either case (00).
The middle column is for those who explained the broad meaning but could not explain retraction in science (10)
and the final column is for those who could do both (11).
Looking across each row in the Treatment group, the mean credibility drops. \color{black}The main finding that can be inferred is \color{black}
that the participants who has better understanding of the term, retraction, at the start of the study, took greater account of information about retraction when they rated credibility of SMNPs. \color{black}This pattern was similar and evident across all three health topics as listed in Table \ref{tab:my-table}\color{black}. 
A \color{black}two\color{black}-way ANOVA test on ratings grouped by different levels of retraction knowledge in the Treatment condition gave $p<0.001$ on all three health topics. \color{black}The ANOVA test table is in the supplementary materials Section~\ref{sec:appex_demographic}\color{black}.
The control group did not show this trend. 
In addition, in line with the consistent picture in Figure~\ref{fig:cred_ratings}, the means for the Treatment group are consistently lower in each topic and knowledge level than the control group.

\begin{table}[h]
\caption{Initial retraction knowledge and credibility ratings}
\label{tab:my-table}
\resizebox{0.9\textwidth}{!}{%
\begin{tabular}{
>{\columncolor[HTML]{FFFFFF}}l 
>{\columncolor[HTML]{FFFFFF}}c 
>{\columncolor[HTML]{FFFFFF}}l 
>{\columncolor[HTML]{FFFFFF}}c 
>{\columncolor[HTML]{FFFFFF}}l 
>{\columncolor[HTML]{FFFFFF}}c }
                                                                  & \multicolumn{5}{c}{\cellcolor[HTML]{FFFFFF}\textbf{Retraction Understanding from RQ1 (General, Science)}}                                                                                                                                                                                                                                                                                   \\ \cline{2-6} 
                                                                  & \multicolumn{1}{c}{\cellcolor[HTML]{FFFFFF}(00)}    &                                              & \multicolumn{1}{c}{\cellcolor[HTML]{FFFFFF}(10)} &                                              & (11)                                                                 \\ \cline{2-2} \cline{4-4} \cline{6-6} 
\textbf{Treatment}                                                & \multicolumn{1}{c}{\cellcolor[HTML]{FFFFFF}P4,7,10,13,28,33,40} &                                              & \multicolumn{1}{l}{\cellcolor[HTML]{FFFFFF}P3,5,17,21,25,32}             &                                              & \multicolumn{1}{l}{\cellcolor[HTML]{FFFFFF}P11,16,19,23,29,35,37,41,44}              \\
\textbf{\begin{tabular}[c]{@{}l@{}}\textit{Masks} -ve\\ \end{tabular}} & 3.6                                                                         & \multicolumn{1}{r}{\cellcolor[HTML]{FFFFFF}} & 2.3                                                                                & \multicolumn{1}{r}{\cellcolor[HTML]{FFFFFF}} & 0.9                                                                                                  \\
\textbf{\textit{Diet}}                                                     & 4.4                                                                         &                                              & 2.9                                                                                &                                              & 1.6                                                                                                  \\
\textbf{\textit{Movie}}                                                    & 4.0                                                                         &                                              & 2.0                                                                                &                                              & 1.6                                                                                                  \\
\textbf{}                                                         & \multicolumn{1}{l}{\cellcolor[HTML]{FFFFFF}}                                &                                              & \multicolumn{1}{l}{\cellcolor[HTML]{FFFFFF}}                                       &                                              & \multicolumn{1}{l}{\cellcolor[HTML]{FFFFFF}}                                                         \\
\textbf{Control}                                                  & \multicolumn{1}{l}{\cellcolor[HTML]{FFFFFF}P6,9,12,14,27,30}      &                                              & \multicolumn{1}{l}{\cellcolor[HTML]{FFFFFF}P2,20,24,42}                      &                                              & \multicolumn{1}{l}{\cellcolor[HTML]{FFFFFF}P1,8,15,18,22,26,31,34,36,38,39,43} \\
\textbf{\begin{tabular}[c]{@{}l@{}}\textit{Masks} -ve\\ \end{tabular}} & 5.3                                                                         & \multicolumn{1}{r}{\cellcolor[HTML]{FFFFFF}} & 4.0                                                                                & \multicolumn{1}{r}{\cellcolor[HTML]{FFFFFF}} & 3.9                                                                                                  \\
\textbf{\textit{Diet}}                                                     & 5.8                                                                         &                                              & 6.7                                                                                &                                              & 6.4                                                                                                  \\
\textbf{\textit{Movie}}                                                    & 5.9                                                                         &                                              & 5.4                                                                                &                                              & 4.9                                                                                                 
\end{tabular}
}
\end{table}

\subsection{RQ4: How does reading SMNP with retraction information change people’s belief on health topics?}
\label{sec:RQ4_impact_prior_knowledge}

\begin{figure}[]
\begin{subfigure}{0.93\textwidth}
  \centering
  \caption{\textit{Mask} pre post belief}
  \includegraphics[width=0.93\linewidth]{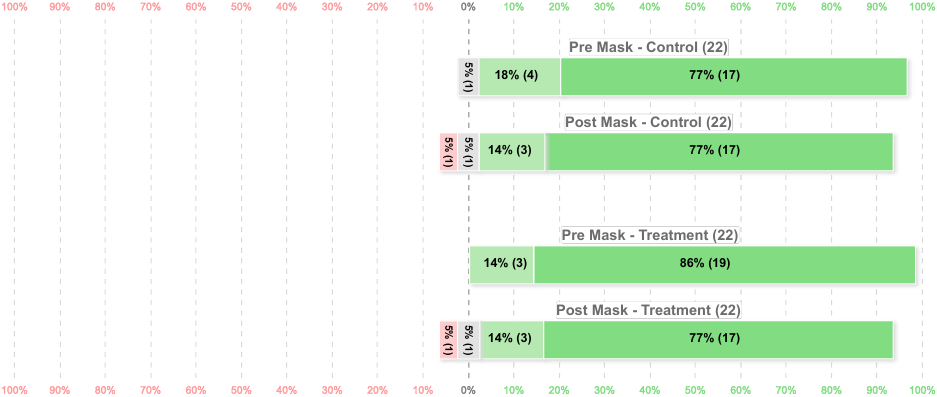}  
  \label{fig:pre_post_mask_stance}
\end{subfigure}
%

\begin{subfigure}{0.93\textwidth}
  \centering
  \caption{\textit{Diet} pre post belief}
  \includegraphics[width=0.93\linewidth]{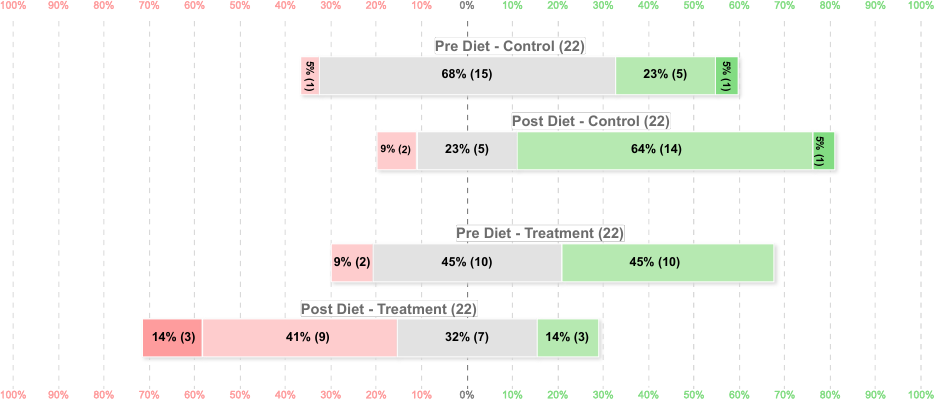}  
  \label{fig:pre_post_diet_stance}
\end{subfigure}

\begin{subfigure}{0.93\textwidth}
  \centering
  \caption{\textit{Movie} pre post belief}
  \includegraphics[width=0.93\linewidth]{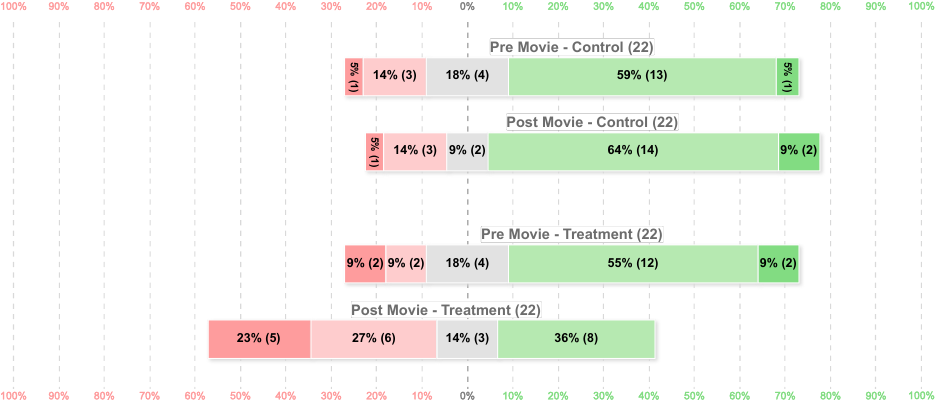}  
  \label{fig:pre_post_movie_stance}
\end{subfigure}

%

\caption{\textbf{Participant belief ratings}. Each pair of  plots shows changes from pre- to post-ratings, showing the Control then Treatment groups for each topic. Colours: Strongly Disagree (dark red) to Neither (grey) to Strongly Agree (dark green). The horizontal position of each bar plot indicates the average of ratings.}
\label{fig:pre_post_stance}
\end{figure}

In Activity \textbf{A3}, in Stage \textbf{S3} in Figure~\ref{fig:study_flowchart}, 
participants rated their agreement with the three health claims.
Figure~\ref{fig:pre_post_stance} summarises participants' pre- and post-test belief responses.
This section describes the trends in the figure and reports representative comments from participants who changed their beliefs.

\subsubsection*{\textbf{Masks Topic}}
The top four bars in Figure~\ref{fig:pre_post_stance}(a) show that most participants maintained their initial strong beliefs on \textit{Masks}. 
Just 11 participants changed their belief, 
four from Control and seven from Treatment. 
In the Control, \color{black}\textbf{P30$_{C_{00}}$} \color{black}
and \color{black}\textbf{P34$_{C_{11}}$} \color{black} 
changed from somewhat (prior) to strongly agree (post), commenting on their previous knowledge that masks are effective.
By contrast, \color{black}\textbf{P6$_{C_{00}}$} \color{black} and \color{black}\textbf{P18$_{C_{11}}$} \color{black} changed from strongly agree to somewhat disagree, commenting on the influence of the SMNPs:\newline 
\noindent

\textit{``I would say disagree, [because I saw] 3 or 4 articles [SMNP] saying that they [masks] are not effective [in limiting spread of COVID-19]."}.(\color{black}\textbf{P6$_{C_{00}}$} \color{black})\vspace{0.2cm}

In the Treatment group, 5 reduced their score with two, \color{black}\textbf{P17$_{T_{10}}$} \color{black} and \color{black}\textbf{P44$_{T_{11}}$} \color{black} 
referring to the influence of the negative \textit{Masks} SMNPs: 
\vspace{0.2cm}

\textit{``now I don't know after reading all that [SMNP], I would say somewhat disagree, all the articles [SMNPs] I have seen [were retracted]."}(\color{black}\textbf{P17$_{T_{10}}$} \color{black}),

\vspace{0.2cm}

The two who increased their score
(from somewhat to strongly agree)
\color{black}\textbf{P35$_{T_{11}}$} 
said this was because the underlying scientific article was not retracted and the SMNP claims were in line with their prior beliefs:  
Broadly, in this topic where participants initially had strong agreement with the claim, there was a little change but those who did change often mentioned the influence of the SMNPs or, in the Treatment group, the retraction information. \color{black}The small change (-0.09 for Control and 0.23 for Treatment) was statistically significant with a p-value<0.01 as shown in section appendix~\ref{sec:belief_change_compare}\color{black} 

\subsubsection*{\textbf{Diet Topic}}
In this case, Figure\ref{fig:pre_post_stance}(b) shows that the Control group, in the first two charts, had more participants move to agreement in the post-test.\color{black} This was also observed in the post-hoc test with coefficients estimate (0.36) with p-values <0.001.\color{black} 
The Treatment group shifted in the opposite direction.
Control group participant comments often indicated the impact of the SMNPs in this topic where they previously had little knowledge, for example:
\vspace{0.2cm}

\textbf{Pre}:\textit{``not sure what Mediterranean diet is, so put that [to] neither"} 
\myarrow[0.5pt]  
\textbf{Post}:\textit{``I strongly agree, since I have not learned about it that much I would say 90\%, I [am] pretty sure but yeah like just like more knowledge can be there"}.
\vspace{0.2cm}

\color{black}\textbf{P12$_{C_{00}}$} \color{black} who initially strongly agreed changed to somewhat agree when she found that the underlying article was retracted and commented \vspace{0.2cm}

\textit{``Somewhat agree because this one had some retracted source in it. but in general I would say still effective}. (\color{black}\textbf{P18$_{C_{11}}$} \color{black}).
\vspace{0.2cm}
 
In the Treatment group, out of 22, 14 participants reduced their previous belief, with 11 of them referring to the retraction status of the article, \color{black}this was visible in the post-hoc test coefficients estimate (-0.91) with p-values <0.001. Notable examples are\color{black}:\vspace{0.2cm}

\textit{``somewhat disagree, because most of the studies [that] have been mentioned [in SMNP], talked about the diet, those studies have been retracted."}(\color{black}\textbf{P10$_{T_{00}}$}\color{black}), \textit{``I am going to say neither, because a lot of the claims I did not find it too credible, although the scientific articles are retracted, so obviously it would reduce it [credibility]"} (\color{black}\textbf{P17$_{T_{10}}$}\color{black})
\vspace{0.2cm}

\subsubsection*{\textbf{Movie Topic}}
Here the Control group, shown in the first two bar plots in Figure~\ref{fig:pre_post_stance}(c) remained similar.\color{black} Post-hoc analysis indicates a small change of 0.136 in the estimated coefficient of post-belief with a p-value <0.01\color{black}.
The Treatment, shifted towards disagreement with comments showing the impact of the retraction information.
Ten participants who reduced their belief also reduced their belief in \textit{Movie} topic. Nine out of them mentioned retraction,\color{black} this was also visible in the post-hoc test coefficients estimate (-0.82) with p-values <0.001. Notable examples are\color{black}: \vspace{0.2cm}

\textit{``I am going to put somewhat disagree because in my mind all I can remember is I am constantly seeing the retracted articles"}(\color{black}\textbf{P41$_{T_{11}}$}\color{black}), \textit{``strongly disagree, 100\%, that sounds completely bogus and all was retracted information from these weird websites."}(\color{black}\textbf{P44$_{T_{11}}$}\color{black})
\vspace{0.2cm}

Personal experience was repeatedly mentioned even in the post-rating $-$
eight participants still agreed with the claim, based on their personal experience. 
For example, \color{black}\textbf{P10$_{T_{00}}$}\color{black} changed from strongly to somewhat agree, saying: 
\textit{``somewhat agree even though some other report [SMNP] shows that it[the study] has been retracted, I still have my knowledge"}.

\color{black}
\subsection*{Participant engagement with SMNP}
We now consider the quantitative data about participant engagement.
This contributes to our understanding of the extent to which the Treatment group actually did make use of the retraction information that was available to them.
This complements the results above.
\color{black}
With the 44 participants equally split across Control and Treatment conditions,
each group saw 264 SMNPs in Stage 3, Activity 2, S3-A2 (12 for each of the 22 participants).
The Treatment group clicked the \textit{More Information} button 240 times (239 clicked, 1 unclicked by mistake).
This means that they checked for more information in 91\% of the SMNPs.
In every one of the 240, participants rated the retraction information 
as useful in determining the credibility of social media science news. 
\textit{Like} and \textit{Share} engagements were much lower in the Treatment condition:
\begin{itemize}
    \item  54 likes and 19 shares in the Treatment condition;
    \item 120 likes and 32 shares in the Control condition. 
\end{itemize}
Participants in the Control group commented that this was their normal behaviour when reading their news feed. 
In contrast, comments from participants in the Treatment group indicated that they wanted to avoid spreading the misinformation.

\color{black}
\subsection*{Belief Change Summary}


The results in Figure~\ref{fig:pre_post_stance} show the impact of prior beliefs on the way that the SMNPs led participants to change their post beliefs.
In the \textit{Masks} topic, where participants were confident in their prior beliefs,
both Control and Treatment groups maintained those beliefs;
they were not influenced by the SMNPs.
In the \textit{Diet}, most participants were not initially aware of the health claim reflected in the large grey bar (showing 68\% neither agree nor disagree).
In this case, the SMNPs had a very different impact:
the Control group increased their belief in the claim
but the Treatment group shifted to disagreeing with it. 
For the \textit{Movie} topic, where participants reported personal experience of eating mindlessly in movies, they initially agreed with the claim.
Here too, we see the impact of the retraction information.
Where the Control group largely maintained their beliefs, the Treatment group shifted towards disagreeing.
Overall, strong prior beliefs (\textit{Masks}) were not altered by the SMNPs, with or without retraction information.
But, in the \textit{Diet} and \textit{Movie} topics, where participants had weaker confidence in their initial beliefs, the retraction information had a strong impact on revised post beliefs. 
\color{black}

\color{black}

\section{Discussion and Implications}
\label{sec:discussion_implcaiton}

\color{black}
We begin with a high level summary of our findings in terms of the broad problem of tackling misinformation due to retracted science. 
There are three main elements to this:
(1) people need to understand scientific retraction;
(2) people need to be able to determine when the SMNP they see is based on retracted science; and
(3) we need the infrastructures that enable SMNP interfaces to quickly and easily access relevant retraction information. 
Our work has been focused on the first two aspects but it also informs the work needed for the third.

For RQ 1 and RQ2 about understanding of retraction key results are:
\begin{itemize}
    \item the study participants were highly educated and had some awareness of the nature of scientific publication;
    \item even so, they had limited understanding of scientific retraction at the start of the study;
    \item most felt they did understand it by the end of the tutorial;
    \item our think-aloud study reveals interesting new insights about participants' perceptions of the impact of each of the seven different reasons for retraction;
\end{itemize}


For RQ3 and RQ4 on the impact of available retraction information:
\begin{itemize}
    \item prior beliefs had a strong impact on the treatment group's assessment of credibility of SMNPs based on retracted articles, and they were familiar only with \textit{Masks};
    \item the Treatment group participants engaged with the retraction information and they changed their beliefs for unfamiliar topics, \textit{Diet} and \textit{Movies}.
    \item our detailed, qualitative study alongside quantitative provides a rich picture of our participants' reasoning about their prior beliefs, how the SMNPs affected their beliefs and how the retraction information played into whether they changed their beliefs or not.
\end{itemize}
Overall, this paints a quite positive picture for the potential to overcome misinformation on SMNPs based on retracted information.
Once people understand the notion of retraction, 
and they can access retraction information, 
this can influence their beliefs on topics where they had limited prior knowledge. We now discuss these findings and their implications in more detail. 
\color{black}

\subsection*{RQ1: How do people understand the retraction of scientific publications?}

Scientific retraction is a quite complex concept that relies on and understanding of the review and publication process.
One third (13 out of 44) of our participants did not even know the general word, retraction.
Half (21) of our participants could not define it in the context of science publication $-$ all of these people could define the general word. 
Given that our participants are university psychology students at a selective university,
we expect the broader population to be even less familiar with the science retraction.
Promisingly, after completing our very simple tutorial, taking around five minutes, most participants felt that they understood scientific retraction and the various reasons for it.
This indicates that schools curricula designed to teach about misinformation could include similar forms of tuition, especially by drawing on our results for RQ2. 

\subsection*{RQ2: What reasons for retractions do people consider to impact the credibility of scientific claims}

Our study gives many insights into participants perceptions of the impact of different reasons for retraction.
Even in the initial questions on defining retraction in scientific contexts, participants referred to aspects like flaws in methodology and analysis.
None mentioned duplicate publishing, plagiarism or permission as reasons for retraction.

\color{black}
The think-aloud comments as participants rated the retractions reasons reveal a consistent picture for some reasons and a much more complex one for others.
We now summarise these results with the most consistent ones first. 
\begin{itemize}
    \item Fabrication, Falsification were seen as reasons to distrust the results (by 43 of 44 participants).
    \item For Duplicate Publishing, there was broad agreement (35 out of 44) that it did \textit{not} affect the validity $-$ but a minority noted that this is not ethical and so brings other aspects into doubt and one fifth felt that getting through the review process twice increases confidence in the results.
    \item Errors were mainly judged to invalidate the results (34 of 44) but some participants highlighted that some forms of errors are common in all science.
    \item Plagiarism had very mixed views $-$ only half  (22) saw it as a reason to doubt results (with 9 unsure).
    \item Reproducibility had half (21 of 44) saying it compromised the results - a surprising result for students how had learnt about the reproducibility crisis in psychology!
    \item Permission had participants split across seeing it compromising results (15), not (17) and being unsure (12).
\end{itemize}
Results were consistent across the control and treatment groups.
Our results match those of Greitemeyer~\cite{greitemeyer2014article}
who only studied retraction due to fabrication.
\color{black}
\color{black}

\subsection*{Implications of informing people about retraction reasons}

Our work indicates that participants saw some reasons as having very different impact from others.
This suggests that it may be useful for interfaces to go beyond simply flagging retraction, and to give the reason as well.
This aligns with work indicating that people can make better credibility judgements
when they see a specific reason (warning) \color{black} with additional context~\cite{sharevski2022meaningful} rather than 
\color{black} 
a generic warning~\cite{gwebu2021can}.
Currently, retracted papers are not marked up with the reasons for retraction.
If they were, then SMNPs could be updated with such retraction information for the many people who consume science news on social media~\cite{hitlin2018science}.
\color{black}Even then, two problems would remain. \color{black}First is the current lack of consensus on ways to report retraction reasons. For example, some journals fail to distinguish scientific error from misconduct~\cite{wager2011and} or they report plagiarism as a significant originality issue~\cite{resnik2013scientific}. 
Potentially, an online tool could automate the COPE guidelines~\cite{cope2019retraction}.  \color{black} A second problem relates to the complex picture we have identified about the ways people see the impact of each reason. 
A path forward may be to focus on reporting the most common reasons or those that are easiest to understand $-$
for example, state whether they are fraudulent or not, along with the reason, e.g., fabrication (fraudulent) or errors (not fraudulent). 
\color{black} 

\subsection*{Summary of changes in beliefs and confidence}

As there is a strong interaction between Research Questions 3 and 4, we now summarise results for both in Table~\ref{tab:summary-belief}.
Bold entries indicate the most important results.
The first pair of columns show SMNP ratings.
The Treatment group has lower ratings of SMNPs for all but the articles supporting \textit{Masks} (where the underlying paper was not retracted).
The lowest credibility score is for the Treatment group on articles refuting \textit{Masks} and the Control group rates it lower than other topics.
The second pair of columns shows changes in participants' agreement with the statements about each health topic. 
Access to retraction information
had a clear impact for both \textit{Diet} and \textit{Movies},
with the Treatment group 
moving from weak agreement to weak disagreement.
For \textit{Masks}, where participants initially felt well informed, this did not happen.
The last set of pair of columns shows that participants became more confident in their beliefs in both conditions and overall for the three topics.

\begin{table}[h]
\centering
\caption{Summary of participant beliefs and credibility ratings. 1) Mean score for SMNP credibility ratings. \textit{Masks} has two scores, the first on supporting \textit{Masks}, and the second for articles refuting \textit{Masks}. 2) Changes in beliefs, by condition. Average initial score .. post-score. 3) Changes in confidence by condition. Mean initial confidence .. post-confidence. Bold indicates key results. }
\label{tab:summary-belief}
\resizebox{0.96\columnwidth}{!}{
\begin{tabular}{llcclcclcc}
\textbf{}       &   & \multicolumn{2}{c}{\textbf{1. SMNP ratings}} &  & \multicolumn{2}{c}{\textbf{2. Change in beliefs}} &  & \multicolumn{2}{c}{\textbf{3. Change in confidence}} \\
Ranges          & = & \multicolumn{2}{c}{{[}0 .. 10{]}}                &  & \multicolumn{2}{c}{{[}-2 .. 2{]}}                 &  & \multicolumn{2}{c}{{[}0 .. 100{]}}                   \\
                &   & \multicolumn{1}{l}{}    & \multicolumn{1}{l}{}   &  & \multicolumn{1}{l}{}    & \multicolumn{1}{l}{}    &  & \multicolumn{1}{l}{}      & \multicolumn{1}{l}{}     \\
\textbf{Topics} &   & \textbf{Control}        & \textbf{Treat}         &  & \textbf{Control}        & \textbf{Treat}          &  & \textbf{Control}          & \textbf{Treat}           \\
Masks           &   & 7.3, 4.3                & 7.6, \textbf{2.1}               &  & 1.7 .. 1.6              & 1.9 .. 1.6              &  & 66 .. 72                  & 75 .. 81                 \\
Diet            &   & 6.3                     & \textbf{3.0}           &  & 0.3 .. 0.6              & \textbf{0.4 .. -0.6}              &  & 24 .. 47                  & 34 .. 45                 \\
Movies          &   & 5.3                     & \textbf{2.6}           &  & 0.5 .. 0.6              & \textbf{0.5 .. -0.4}    &  & 46 .. 61                  & 45 .. 55                
\end{tabular}
}
\end{table}

\subsection*{RQ3: How does a person’s prior belief affect their rating of the credibility of SMNP, that has information about retraction?}
\color{black} 
Overall, Treatment group participants appear to have been strongly influenced by access to information about retraction.
Our careful selection of health topics provided insights about the impact of different levels of prior knowledge. \color{black}
In the case of \textbf{Masks}, where they already felt well informed, they discounted the impact of retraction for the article supporting \textit{Masks}. 
At the same time, we saw a move to lower ratings in the Control group for articles negating value with \textit{Masks} and based on retracted scientific work.
This matches the literature showing people give greater credibility to articles confirming their prior beliefs compared to those refuting them.
~\cite{flynn2017nature,pennycook2021psyschology,shahid2022matches}. 

For the \textbf{Diet} SMNPs, where many participants (25 out of 44) initially felt uninformed and had low confidence in their rating,
we see a large impact of the availability of retraction information.
We see a similar overall effect in the \textbf{Movie} SMNP ratings, even though participants had personal experience that was consistent with the claims in the SMNP.
Indeed, some participants continued to rely on that experience, in line with observations that people discount information that clashes with their worldview~\cite{flynn2017nature, fogg1999elements}.  
Regardless of prior belief in all three topics, participants who saw the retraction through the interface rated the SMNP on average at least two units (on a scale of 0..10) lower than those in the respective control condition. \color{black}
This is in line with the study by Greitemeyer~\cite{greitemeyer2014article} where participants were also Psychology students.
The debriefing group had lower belief in the retracted study findings compared to the no-debriefing group. \color{black}

Participants who could show understanding of  
retraction in the first stage of the study
rated the SMNPs credibility lower than the other participants.
Even those participants who \textit{only} understood the general meaning of the word also rated the SMNP lower than those who did not know it. 
This indicates the impact of understanding retraction on people's ability to make use of information about it. 
This points to the potential impact of education about this topic.

\color{black}
We analysed engagement as an measure of how much Treatment participants actually used the retraction information and how that was reflected in their actions on it.
Treatment group participants shared and liked the posts half as much as Control participants. This is in line with the research that suggests adding lightweight interventions reduces sharing intent~\cite{yaqub2020effects}. 
Our participants sharing intentions of true news or non-retracted based news were not impacted by our retraction information. 
This is unlike what was found by the lightweight intervention in~\cite{jahanbakhsh2021exploring} where true news sharing was impacted when users were nudged to think whether SMNP headlines were accurate. 
Our participants had similar political inclinations therefore the impact of political ideology was not observed in our findings which was reported by~\cite{yaqub2020effects,lees2021twitter}.

Our results on the impact of the Treatment group's heavy use of the ``More Information'' button
is similar to what Clayton et al.~\cite{clayton2020real} found when adding ``Rated false'' on the disputed flags. On the other hand, perceived credibility increased for non-retracted ones, indicating that an article's retraction status served as a significant factor in their decision-making process. This finding is in line with the implied truth effect observed that non-discredited news is seen as more trustworthy news~\cite{pennycook2020implied}.


One key difference that stands out between warnings or disputed flags created by Fact-checkers on SMNPs in contrast to Retraction-based warnings by Journal editors is that in general people trust scientists and the scientific community~\cite{funk2020science}  which cannot be said about Fact-checkers~\cite{park2021presence}.

\color{black}

\subsection*{RQ4: How does reading SMNP change people’s belief of health topics?}

Table~\ref{tab:summary-belief} shows the strong impact of availability to information on retraction on people's beliefs.
This is promising as it indicates that interfaces like ours may well be part of a solution to this information based on retracted scientific news.

\subsection*{Implications for creating SMNP interfaces that help people}
Participants accessed ``More Information'' in 90.1\% (240 of 264) of the SMNPs and found it useful in determining the credibility of the SMNP each time unlike the current Facebook SMNP interface, where most participants ignored more information ~\cite{geeng2020fake}. Adding ``More information'' near the interaction panel can help users use it more, which was also part of the interface design in the Fakey game~\cite{avram2020exposure}.

\color{black}
\subsection*{Implications for broader populations and in-the-wild reading of SMNPs}

Our results are stronger than one would expect for a broader population and for news in the wild.
This is because of both the population we studied and the artificial context of our study, both as a lab study where people were aware they were being observed and because the first part of our study primed the participants to focus on retraction.
In addition, participants saw high proportion of SMNPs based on retracted science.
Previous work has shown that the laboratory studies like ours can carry over to actual social media behaviour~\cite{mosleh2020self}, 

Our participants were psychology students.
This has important implications for the generalizability of our findings.
For RQ1 and RQ2, which related to knowledge about retraction and its impact on scientific validity,
this population is likely to be far more knowledgeable about the publication process and retraction than the general public.
While our short tutorial was enough for most of them to feel confident about the meaning of retraction in science, 
we would expect that more time would be needed for a broader population.
Even so, our work provides valuable foundations for that instruction as it highlights the complexity of retraction and people's perceptions of the impact of different reasons for it.

For RQ3 and RQ4 on the impact of available retraction information, our results are promising in that they do show that our participants in the lab setting were able to use the information and were influenced by it.
However, translating this to authentic settings can be expected to have lower impact even if people reading SMNPs do know about retraction.
There is much to be learnt about how much people will engage with retraction information.
There are also many possible avenues to explore as ways to help people become aware of their use of available information and to encourage its use.

\color{black}

\subsection*{Limitations and future work} \label{sec:limitations}

Our study was limited to 44 participants,
all first- and second-year psychology students at the University of Anon.
As discussed above, the nature of the population and the artificial setting, provide a valuable baseline of knowledge about this area. 
\color{black}Exposure to retraction and its reasons as a common knowledge base through tutorial prior to credibility assessments could indicate harsher ratings in Treatment group \color{black}.
It would be valuable to replicate this study with a broader population in a more authentic setting such as a crowd-sourced study, or better yet, with social media sites tracking authentic user activity when retraction information is available. 
Our headlines are all on health topics and our findings may not carry over to more polarising topics.
It would be valuable to replicate this work for such topics.


\section{Conclusion}
\label{sec:conclusion}

To tackle the problems of misinformation due to retracted science, we need to tackle three problems:
\begin{itemize}
    \item people need to understand scientific retraction and its implications;
    \item people reading SMNPs need to have information about such retraction and take account of it;
    \item we need the infrastructures that can power the delivery of this information.
\end{itemize}
Our work has tackled the first two of these and the findings about the importance of the reasons for retraction have implications for the last.

Our mixed methods study provides an rich, in-depth understanding of our participant's knowledge of retraction in general, and in a scientific publication context.
It also gives insights about the way they perceived the impact of different reasons for retraction on the credibility of its claims.
It also gives foundations for teaching about retraction.

We designed the study with three health topics,
each with a different level of familiarity for participants,
based on both their knowledge and their personal experience.
This revealed a nuanced set of responses in the ratings of the SNMPs
and in the changes in belief, and certainty about that belief, across the topics.
Our work provides new understanding of a little studied but important source of misinformation in SNMPs based on retracted scientific work. 
\bibliographystyle{ACM-Reference-Format}
\bibliography{00_main.bib}

\appendix
\clearpage
\section{\texorpdfstring{\centering{\Large{Stepwise Regression output}}}{Lg}}
\label{sec:stepwise_regression}

\tiny
\begin{lstlisting}[float=h,frame=tb,caption=R output,label=zebra]
filtered_data <- filter(data, health_topic!="tutorial")
**************************************** STEP 1 (Condition variable) **************************************
> anova(model_s0,model_s1)
Data: filtered_data
Models:
model_s0: ratings ~ 1 + (1 | PID) + (1 | QID)
model_s1: ratings ~ Group + (1 | PID) + (1 | QID)
         npar    AIC    BIC  logLik deviance  Chisq Df Pr(>Chisq)    
model_s0    4 2219.1 2236.2 -1105.5   2211.1                         
model_s1    5 2194.7 2216.0 -1092.3   2184.7 26.405  1  2.768e-07 ***
---
Signif. codes:  0 `***' 0.001 `**' 0.01 `*' 0.05 `,' 0.1 ` ' 1
**************************************** STEP 2 (Health topic)**************************************

> anova(model_s1,model_s2)
Data: filtered_data
Models:
model_s1: ratings ~ Group + (1 | PID) + (1 | QID)
model_s2: ratings ~ Group * health_topic + (1 | PID) + (1 | QID)
         npar    AIC    BIC  logLik deviance  Chisq Df Pr(>Chisq)    
model_s1    5 2194.7 2216.0 -1092.3   2184.7                         
model_s2   11 2093.9 2140.8 -1036.0   2071.9 112.79  6  < 2.2e-16 ***

---
Signif. codes:  0 `***' 0.001 `**' 0.01 `*' 0.05 `,' 0.1 ` ' 1


**************************************** STEP 3 (Prior Beliefs) **************************************
> anova(model_s2,model_s3)
Data: filtered_data
Models:
model_s2: ratings ~ Group * health_topic + (1 | PID) + (1 | QID)
model_s3: ratings ~ Group:health_topic:as.factor(Pre.Masks) + 
Group:health_topic:as.factor(Pre.Movie) + Group:health_topic:as.factor(Pre.Diet) + 
(1 | PID) + (1 | QID)
         npar    AIC    BIC   logLik deviance  Chisq Df Pr(>Chisq)    
model_s2   11 2093.9 2140.8 -1035.96   2071.9                         
model_s3   71 2076.8 2379.6  -967.43   1934.8 137.06 60  5.804e-08 ***


*********************************** Summary of the final model ************************************

summary(model_s3)

Fixed effects:
                                               Estimate Std. Error    df     t value Pr(>|t|) 
GroupControl  :mask_positive:(Pre.Masks.)0     4.48678    2.55888  98.66824   1.753 0.082636 .  
GroupControl  :mask_negative:(Pre.Masks.)1     0.82295    2.53181 121.05105   0.325 0.745708    
GroupTreatment:mask_negative:(Pre.Masks.)1    -0.82717    1.66870 459.68961  -0.496 0.620342    
GroupControl  :mask_positive:(Pre.Masks.)1     7.59086    2.42757 109.31881   3.127 0.002264 ** 
GroupTreatment:mask_positive:(Pre.Masks.)1     5.61482    2.28762 511.15186   2.454 0.014443 *  
GroupControl  :mask_negative:(Pre.Masks.)2     1.62230    2.18261 107.22224   0.743 0.458937    
GroupTreatment:mask_negative:(Pre.Masks.)2    -1.02769    1.38484 505.47613  -0.742 0.458373    
GroupControl  :mask_positive:(Pre.Masks.)2     5.66312    2.20478 104.25038   2.569 0.011628 *  
GroupTreatment:mask_positive:(Pre.Masks.)2     4.75935    1.96473 513.17122   2.422 0.015764 *  
GroupControl  :movie_positive:(Pre.Movie.)-1   0.23216    1.26672  77.37862   0.183 0.855062    
GroupTreatment:movie_positive:(Pre.Movie.)-1   0.95635    1.19355  78.26400   0.801 0.425404    
GroupControl  :movie_positive:(Pre.Movie.)0    0.53233    1.22281  78.27496   0.435 0.664518    
GroupTreatment:movie_positive:(Pre.Movie.)0    2.10072    0.96006  78.60169   2.188 0.031632 *  
GroupControl  :movie_positive:(Pre.Movie.)1    1.26089    1.18881  77.74564   1.061 0.292142    
GroupTreatment:movie_positive:(Pre.Movie.)1    0.29480    0.83430  77.72172   0.353 0.724786    
GroupControl  :movie_positive:(Pre.Movie.)2    2.02504    1.66184  77.42938   1.219 0.226714    
GroupTreatment:movie_positive:(Pre.Movie.)2    3.43199    1.15509  78.65341   2.971 0.003934 ** 
GroupControl  :diet_positive:(Pre.Diet.)0     -0.49089    1.19065  78.22572  -0.412 0.681256    
GroupTreatment:diet_positive:(Pre.Diet.)0      0.79980    0.93132  80.71390   0.859 0.393005      
GroupControl  :diet_positive:(Pre.Diet.)1      1.50639    1.24664  77.63963   1.208 0.230576    
GroupTreatment:diet_positive:(Pre.Diet.)1      1.13449    1.00436  79.64641   1.130 0.262047    
GroupControl  :diet_positive:(Pre.Diet.)2     -0.28943    1.53408  78.51812  -0.189 0.850840     
\end{lstlisting}
\normalsize

\clearpage
\section{\texorpdfstring{\centering{\Large{Pairwise comparison of subgroups Group and Health topic}}}{Lg}}
\label{sec:subgroup_compare}

We further conducted post-hoc tests and calculated estimated marginal means in linear mixed effects models to see if there are statistically significant differences between treatment and control divided by health topic. To compare subgroups in our lmer model, we used emmeans package with Bonferroni adjustments to calculate the estimated marginal means and perform pairwise comparisons with multiple comparisons adjustment.

\tiny
\begin{lstlisting}[float=h,frame=tb,caption=R output,label=zebra]
 contrast                                                estimate    SE    df t.ratio p.value
 Control   diet_positive -  Treatment diet_positive      3.321 0.436  72.4   7.614  <.0001  ---|
 Control   movie_positive - Treatment movie_positive     2.627 0.438  73.5   5.995  <.0001     | control  
 Control   mask_negative -  Treatment mask_negative      1.873 0.508 125.3   3.690  0.0093     | vs Treatment
 Control   mask_positive -  Treatment mask_positive      0.270 0.528 142.2   0.512  1.0000  ---| ratings
 Control   diet_positive -  Control mask_negative        2.049 0.362 105.0   5.659  <.0001
 Control   diet_positive -  Treatment mask_negative      3.922 0.499  91.7   7.864  <.0001
 Control   diet_positive -  Control mask_positive       -1.057 0.348  82.0  -3.038  0.0895
 Control   diet_positive -  Treatment mask_positive     -0.787 0.529 112.8  -1.489  1.0000
 Control   diet_positive -  Control movie_positive       1.058 0.301  54.7   3.512  0.0252
 Control   diet_positive -  Treatment movie_positive     3.685 0.472  75.0   7.807  <.0001
 Treatment diet_positive -  Control mask_negative       -1.272 0.514 102.8  -2.476  0.4182
 Treatment diet_positive -  Treatment mask_negative      0.602 0.341  84.8   1.767  1.0000
 Treatment diet_positive -  Control mask_positive       -4.378 0.504  93.5  -8.689  <.0001
 Treatment diet_positive -  Treatment mask_positive     -4.108 0.383 126.2 -10.712  <.0001
 Treatment diet_positive -  Control movie_positive      -2.262 0.473  75.4  -4.786  0.0002
 Treatment diet_positive -  Treatment movie_positive     0.365 0.300  55.0   1.215  1.0000
 Control   mask_negative -  Control mask_positive       -3.106 0.416 145.5  -7.461  <.0001
 Control   mask_negative -  Treatment mask_positive     -2.836 0.567 142.1  -5.006  <.0001
 Control   mask_negative -  Control movie_positive      -0.991 0.363 105.3  -2.726  0.2102
 Control   mask_negative -  Treatment movie_positive     1.636 0.514 102.6   3.185  0.0536
 Treatment mask_negative -  Control mask_positive       -4.980 0.529 110.5  -9.421  <.0001
 Treatment mask_negative -  Treatment mask_positive     -4.709 0.437 181.1 -10.775  <.0001
 Treatment mask_negative -  Control movie_positive      -2.864 0.499  91.9  -5.736  <.0001
 Treatment mask_negative -  Treatment movie_positive    -0.237 0.341  83.8  -0.695  1.0000
 Control   mask_positive -  Control movie_positive       2.116 0.349  81.8   6.068  <.0001
 Control   mask_positive -  Treatment movie_positive     4.743 0.504  93.2   9.413  <.0001
 Treatment mask_positive -  Control movie_positive       1.845 0.530 113.0   3.484  0.0197
 Treatment mask_positive -  Treatment movie_positive     4.473 0.383 125.0  11.667  <.0001
 

Degrees-of-freedom method: kenward-roger 
P value adjustment: Bonferroni method
\end{lstlisting}
\normalsize

\clearpage

\section{\texorpdfstring{\centering{\Large{Pairwise comparison of subgroups Group and Health topic}}}{Lg}}
\label{sec:belief_change_compare}

We further conducted post-hoc tests and calculated estimated marginal means in linear mixed effects models to see if there are statistically significant differences between post belief of participants in treatment and control divided by health topic. To compare subgroups in our lmer model, we used emmeans package with Bonferroni adjustments to calculate the estimated marginal means and perform pairwise comparisons with multiple comparisons adjustment. The Bonferroni adjustment did not impact the p-values because the subgroups were one-to-one group comparisons.

\tiny
\begin{lstlisting}[float=h,frame=tb,caption=R output,label=zebra]
Group = Control:
 contrast                                             estimate     SE   df t.ratio p.value
 Post.Diet - Pre.Diet                                 0.364 0.0396 1010   9.174  <.0001

Group = Treatment:
 contrast                                             estimate     SE   df t.ratio p.value
 Post.Diet - Pre.Diet                                -0.909 0.0396 1010 -22.936  <.0001
 
 Group = Control:
 contrast                                             estimate     SE   df t.ratio p.value
 Post.Movie - Pre.Movie                                 0.136 0.0474 1010   2.879  0.0041

Group = Treatment:
 contrast                                             estimate     SE   df t.ratio p.value
 Post.Movie - Pre.Movie                                -0.818 0.0474 1010 -17.271  <.0001
 
 Group = Control:
 contrast                                             estimate     SE   df t.ratio p.value
 Post.Masks - Pre.Masks                               -0.0909 0.0326 1010  -2.785  0.0055

Group = Treatment:
 contrast                                             estimate     SE   df t.ratio p.value
 Post.Masks - Pre.Masks                               -0.2273 0.0326 1010  -6.962  <.0001
\end{lstlisting}
\normalsize

\clearpage
\section{\texorpdfstring{\centering{\Large{Supplementary Materials}}}{Lg}}
\label{sec:supp_materials}

\subsection{\texorpdfstring{\centering{\Large{Tutorial and underlying details for each retraction reasons}}}{Lg}} \label{sec:appex_retract_tutorial}

This section lists the full text of the information available to participants during the tutorial.

Much of the science news reported is based on valid scientific studies, while some science news is based on scientific studies that are later retracted.\newline\newline
In science, retraction indicates a published article was withdrawn from publication. 
Retraction can occur due to many reasons but frequently due to research misconduct that \textbf{can invalidate scientific study claims}.\newline\newline

\begin{enumerate}
\item Drag the reasons for retraction items in red below to any of the \textbf{three (G1, G2, G3)} appropriate groups. Hover over any item for details.\newline\newline
    \begin{itemize}
        \item Fabrication \textit{(underlying hovered text: Making up data or results rather than having them come from actual research, and recording or reporting them.)}
        \item Duplicate publishing \textit{(underlying hovered text: Being submitted and accepted in more than one publication)}
        \item Falsification \textit{(underlying hovered text: Manipulating research materials, equipment, or processes; changing or omitting data; providing results where the research is not accurately represented)}
        \item Plagiarism \textit{(underlying hovered text: Using another person's ideas, processes, results, or words without giving attribution.)}
        \item Errors \textit{(underlying hovered text: Errors in the research.)}
        \item Reproducibility \textit{(underlying hovered text: Problems with its reproducibility.)}
        \item Permission \textit{(underlying hovered text: Not obtaining proper permissions to use data.)}
    \end{itemize}
    \vspace{0.25cm}
\textit{Groups:} \textbf{G1: Does invalidates the scientific claim
}= [\ldots ] 
    \textbf{G2: Does not invalidates the scientific claim
} = [\ldots ]
    \textbf{Not sure} = [\ldots ]

\vspace{0.25cm}

\end{enumerate}
\clearpage
\subsection{\texorpdfstring{\centering{\Large{Figures and Tables:}}}{Lg}}
\label{sec:appendix}

\begin{figure}[H]
     \centering
     \begin{subfigure}[b]{0.27\textwidth}
         \centering
         \includegraphics[width=\textwidth]{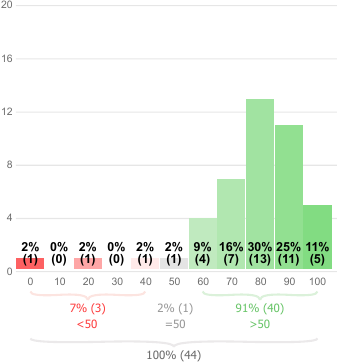}
         \caption{Post mask}
         \label{fig:pre_mask_conf}
     \end{subfigure}
     \hfill
     \begin{subfigure}[b]{0.27\textwidth}
         \centering
         \includegraphics[width=\textwidth]{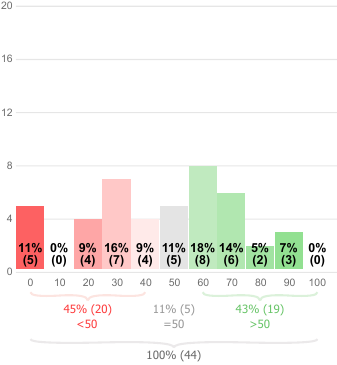}
         \caption{Post diet}
         \label{fig:pre_diet_conf}
     \end{subfigure}
     \hfill
     \begin{subfigure}[b]{0.27\textwidth}
         \centering
         \includegraphics[width=\textwidth]{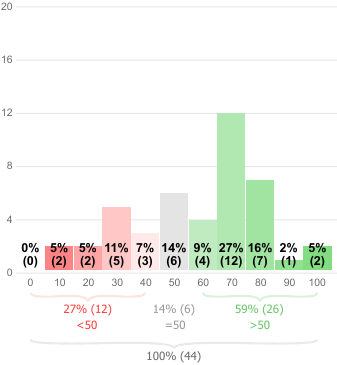}
         \caption{Post movie}
         \label{fig:pre_movie_conf}
     \end{subfigure}
     
             \caption{Confidence distribution reported on the three health topics in the posttest}
        \label{fig:post_three_health_confidence}
\end{figure}


\subsection{\texorpdfstring{\centering{\Large{Summative table for Figure ~\ref{fig:tutorial_retract_response}}}}{Lg}}

\begin{table}[H]
\centering
\caption{Participants responses whether the listed reason invalidates the findings of the paper first three are fraudulent and the rest are non fraudulent reasons for retraction}
\label{tab:summative_table_retract}
\begin{tabular}{lrlrlrllr}
                              & \multicolumn{1}{l}{\textbf{G1}} &  & \multicolumn{1}{l}{\textbf{G2}} &  & \multicolumn{1}{l}{\textbf{G3}} &  &  & \multicolumn{1}{l}{} \\ \cline{2-2} \cline{4-4} \cline{6-6}
\textbf{Fabrication}          & 43                              &  & 1                               &  & 0                               &  &  &                      \\
\textbf{Falsification}        & 43                              &  & 0                               &  & 1                               &  &  &                      \\
\textbf{Plagiarism}           & 22                              &  & 13                              &  & 9                               &  &  &                      \\
\textbf{Errors}               & 34                              &  & 3                               &  & 7                               &  &  &                      \\
\textbf{Reproducibility}      & 21                              &  & 9                               &  & 14                              &  &  &                      \\
\textbf{Permission}           & 15                              &  & 17                              &  & 12                              &  &  &                      \\
\textbf{Duplicate publishing} & 0                               &  & 35                              &  & 9                               &  &  &                      \\
\textbf{}                     &                                 &  &                                 &  &                                 &  &  &                      \\
\textbf{}                     &                                 &  &                                 &  &                                 &  &  &                      \\
\textbf{}                     &                                 &  &                                 &  &                                 &  &  &                     
\end{tabular}
\end{table}

\subsection{\texorpdfstring{\centering{\Large{ANOVA test Table:}}}{Lg}}

\begin{table}[h]
\centering
\caption{ANOVA test table showing }
\label{tab:anova_retract_cred}
\begin{tabular}{lrlrlrlr}
\textbf{Anova Table (Type III tests)} & \multicolumn{1}{l}{}                &                      & \multicolumn{1}{l}{}            &                      & \multicolumn{1}{l}{}                 &                      & \multicolumn{1}{l}{}                             \\
\textbf{}                             & \multicolumn{1}{l}{}                &                      & \multicolumn{1}{l}{}            &                      & \multicolumn{1}{l}{}                 &                      & \multicolumn{1}{l}{}                             \\
\textbf{}                             & \multicolumn{1}{c}{\textbf{Sum Sq}} &                      & \multicolumn{1}{c}{\textbf{Df}} &                      & \multicolumn{1}{c}{\textbf{F value}} &                      & \multicolumn{1}{c}{\textbf{Pr(\textgreater{}F)}} \\ \cline{2-2} \cline{4-4} \cline{6-6} \cline{8-8} 
\textbf{(Intercept)}                  & 1128.76                             & \multicolumn{1}{r}{} & 1                               & \multicolumn{1}{r}{} & 296.3269                             & \multicolumn{1}{r}{} & \textless 2.2e-16 ***                            \\
\textbf{Retract.knowledge}            & 11.22                               & \multicolumn{1}{r}{} & 2                               & \multicolumn{1}{r}{} & 1.4728                               & \multicolumn{1}{r}{} & 0.2309764                                        \\
\textbf{Group}                        & 244.01                              & \multicolumn{1}{r}{} & 1                               & \multicolumn{1}{r}{} & 64.0587                              & \multicolumn{1}{r}{} & 2.902e-14 ***                                    \\
\textbf{Retract.knowledge:Group}      & 57.82                               & \multicolumn{1}{r}{} & 2                               & \multicolumn{1}{r}{} & 7.5890                               & \multicolumn{1}{r}{} & 0.0006126 ***                                    \\
\textbf{Residuals}                    & \multicolumn{1}{l}{1108.47}         &                      & \multicolumn{1}{l}{291}         &                      & \multicolumn{1}{l}{}                 &                      & \multicolumn{1}{l}{}                            
\end{tabular}
\end{table}

\clearpage

\subsection{\texorpdfstring{\centering{\Large{Demographics and Social Media Use}}}{Lg}}
\label{sec:appex_demographic}

The demographic information collected was age, gender, geographic location and highest educational level.
These are important for characterising the population.
We included a question on health literacy since literature showed that people with lower literacy may be more prone to believe misinformation~\cite{damian2020promoting}, notably COVID-19 misinformation~\cite{bin2021covid, wojtowicz2020addressing}.
We asked the question
\cite{cornett2009assessing}:
\textit{How often do you need to have someone help you when you read instructions, pamphlets, or other written material from your doctor or pharmacist?}
Using thresholds from~\cite{morris2006single}
we mapped \textit{always}, \textit{often}, and \textit{sometimes} to low-literacy; 
the answers \textit{occasionally} and \textit{never} are not low literacy.

We asked if participants used social media and for those who did, we asked about the platforms used, the length of use in years.
We asked how much time participants spent reading, watching, or listening to the news on social media.
Details of use are important because we had designed the interfaces to be familiar to people who do consume news on social media. 
We asked participants to rate the news sources that they most trusted and the ones they most used.
These questions, also in 
\cite{yaqub2020effects}
provide descriptive data about our participants.
and for our research questions about participants' assessment of credibility of social media news items.



\begin{enumerate}
    \vspace{0.25cm}
    \item What is your age?
    \begin{multicols}{5}
    \begin{itemize}
        \item 18-25
        \item 26-35
        \item 36-45
        \item 45-50
        \item 60+
    \end{itemize}
    \end{multicols}
    
    \vspace{0.25cm}
    \item Which gender do you identify with the most?
    \begin{multicols}{4}
    \begin{itemize}
        \item Male
        \item Female
        \item Prefer not to say
        \item Other. Please Specify:
    \end{itemize}
    \end{multicols}
    
    \vspace{0.25cm}
    \item Please provide the country, city that you work/live in and number of years.
    \begin{itemize}
        \item Country: 
        \item City:
        \item Years lived: 
    \end{itemize}
    
    \vspace{0.25cm}
    \item What is the highest level of education you have completed? (If currently enrolled, highest degree received.)
    \begin{itemize}
        \item Less than high school
        \item High school graduate
        \item College graduate (B.S., B.A., or other 4 year degree)
        \item Higher Degree
        \item Professional degree after college (e.g., law or medical school)
        \item Prefer not to say
       \item Other. Please Specify:
    \end{itemize}
    
    \vspace{0.25cm}
    \item On a scale of 1-7, with 1 being Extremely Liberal and 7 being Extremely Conservative , where would you place yourself?
    \begin{itemize}
        \item 1 (Extremely Liberal)
        \item 2
        \item 3
        \item 4
        \item 5
        \item 6 
        \item 7 (Extremely Conservative)
    \end{itemize}

    \vspace{0.25cm}
    \item Do you have or had a social media account?
    \begin{itemize}
        \item Yes, currently I have social media account
        \item Yes, I had, but not currently active
        \item No
    \end{itemize}
    
    \vspace{0.25cm}
    \item How many years have you used social media?
        \begin{itemize}
        \item Less than a year
        \item 1 
        \item 2
        \item 3
        \item 4
        \item 5
        \item More than 5
    \end{itemize}

    \vspace{0.25cm}
    \item How often do you use social media?
    \begin{multicols}{5}
    \begin{itemize}
        \item Very often
        \item Often 
        \item Sometimes
        \item Hardly ever
        \item Never
    \end{itemize}
    \end{multicols}
    
    \vspace{0.25cm}
    \item On average, how many minutes do you spend each day reading, watching, or listening to the news on social media?
    
    \vspace{0.25cm}
    \item Which social media platforms do you use? (Group the platforms based on usage, by dragging red platform names to either most frequent, least frequent or don't use group)
    \begin{itemize}
        \item Facebook
        \item Twitter
        \item Instagram
        \item Reddit
        \item Snapchat
        \item Tumblr
        \item LinkedIn
        \item Pinterest
        \item Other. Please specify:
    \end{itemize}
    \vspace{0.25cm}
    \hspace{3cm} \textit{Groups:} \textbf{Most frequent}= [\ldots ] 
    \textbf{Least frequent} = [\ldots ]
    \textbf{Don't use} = [\ldots ]
    
    \vspace{0.25cm}
    \item Which of the following do you consider the most trusted source for news and information?
    \begin{multicols}{3}
    \begin{itemize}
        \item Print newspapers
        \item National television
        \item Web Blogs
        \item Radio
        \item Cable television
        \item Social media (e.g., Facebook, Twitter, Whatsapp)
        \item Local television
        \item Online news Web sites or Apps excluding social media and blogs
        \item Other. Please specify:
    \end{itemize}
    \end{multicols}

    \vspace{0.25cm}
    \item Which of the following sources do you use to get most news and information?
    \begin{multicols}{3}
    \begin{itemize}
        \item Print newspapers
        \item National television
        \item Web Blogs
        \item Radio
        \item Cable television
        \item Social media (e.g., Facebook, Twitter, Whatsapp)
        \item Local television
        \item Online news Web sites or Apps excluding social media and blogs
        \item Other. Please specify:
    \end{itemize}
    \end{multicols}
    
\end{enumerate}

\clearpage

\end{document}